\documentclass{aa}  
\usepackage{graphicx}
\usepackage{float}
\usepackage{txfonts}
\usepackage{color}
\usepackage{float}

\begin{document}

   \title{Tip of the red giant branch distances to the dwarf galaxies dw1335-29 and dw1340-30 in the Centaurus group\thanks{{Based on observations collected at the European Organisation for Astronomical Research in the Southern Hemisphere 
under ESO programme 097.B-0306(A).}}}

   \author{Oliver M\"uller
          \inst{1}
          \and
          Marina Rejkuba\inst{2}
          \and
          Helmut Jerjen\inst{3}
          }

   \institute{Departement Physik, Universit\"at Basel, 
 Klingelbergstr. 82, CH-4056 Basel, Switzerland\\
              \email{oliver89.mueller@unibas.ch}
         \and
         European Southern Observatory, Karl-Schwarzschild Strasse 2, 85748, Garching, Germany
         \and
			Research School of Astronomy and Astrophysics, Australian National University, Canberra,
ACT 2611, Australia
             }

   \date{\today}

 
  \abstract
   {The abundance and spatial distribution of dwarf galaxies are excellent empirical benchmarks to test models of structure formation on small scales. The nearby Centaurus group, with its two subgroups centered on Cen\,A and M\,83, stands out as an important alternative to the Local Group for scrutinizing cosmological model predictions in a group of galaxies context.}
   {We have obtained deep optical images of three recently discovered M\,83 satellite galaxy candidates with the FORS2 instrument mounted on the Very Large Telescope. We aim to confirm their group membership and study their stellar population.}
   {Deep $VI$-band photometry is used to resolve the brightest stars in our targets. Artificial star tests are performed to estimate the completeness and uncertainties of the photometry. The color-magnitude diagrams reveal the red giant branch (RGB) stars allowing to use the Sobel edge detection method to measure the magnitude  of the RGB tip and thus derive distances and group membership for our targets. The mean metallicity of the dwarf galaxies are further determined by fitting BASTI model isochrones to the mean RGB locus. }  
   {We confirm the two candidates, dw1335-29 and dw1340-30, to be dwarf satellites of the M\,83 subgroup, with estimated distances of $5.03\pm0.24$\,Mpc and $5.06\pm 0.24$\,Mpc, respectively. Their respective mean metallicities of $\langle$[Fe/H]$\rangle=-1.79\pm0.4$ and $\langle$[Fe/H]$\rangle=-2.27\pm0.4$ are consistent with the metallicity-luminosity relation for dwarf galaxies. The third candidate, dw1325-33, could not be resolved into stars due to
   insufficiently deep images, implying its distance must be larger than 5.3\,Mpc.
   Using the two newly derived distances we assess the spatial distribution of the galaxies in the M\,83 subgroup {and discuss a potential plane-of-satellites around M\,83}.}
{}


\titlerunning{TRGB distances to dw1335-29 and dw1340-30}
\authorrunning{M\"uller, Rejkuba \& Jerjen}
   \maketitle
%

\section{Introduction}
It is well-known that dwarf galaxies trace the ``fine structure of large-scale structure'' \citep{1989ASSL..151...47B}.
The imprinted information in the spatial distribution of dwarf galaxies provides an excellent test bed for structure formation models. 
Hence the search for optically elusive dwarf galaxies in the Galactic neighborhood is an important contribution to better understand the 3D-distribution of baryonic and non-baryonic matter. Besides using 3D spatial information for cosmology, 
the census of dwarf galaxies in group environments gives insight into the properties of the local universe \citep{1988AJ.....96...73T}.  The need for such fundamental empirical input makes it necessary to systematically search the sky for yet undetected dwarf galaxies. Such an investigation is typically conducted in two steps: (1) find unresolved dwarf candidates in large-field imaging surveys (e.g. \citealp{2001A&A...377..801H,2009AJ....137.3009C,2014ApJ...787L..37M,2015A&A...583A..79M,2015AstBu..70..379K,2016A&A...588A..89J,2017A&A...597A...7M,2017A&A...602A.119M,2017ApJ...848...19P,2017A&A...603A..18H}); and then (2) establish distances with ground or space based telescopes (e.g. \citealp{2001A&A...380...90J,2001A&A...371..487J,2007AJ....133..504K,2013AJ....146..126C,2017ApJ...837..136D,2017ApJ...843L...6S,2018MNRAS.474.3221M}). Alternatively, deep imaging surveys can resolve dwarf galaxies directly into individual stars, especially in the Local Group \citep{2015ApJ...805..130K,2015ApJ...804L..44K} and nearby groups \citep[e.g.][]{2014ApJ...793L...7S, 2014ApJ...780..179M, 2014ApJ...795L..35C,2016ApJ...823...19C}. The latter approach, however, can only be done for nearby dwarfs or by a trade-off in the area of the surveyed field.

The nearby Centaurus Group is made up of the Centaurus\,A (Cen\,A) subgroup at a mean distance of 3.8\,Mpc and the slightly more distant M\,83 subgroup at 4.9\,Mpc \citep{2002A&A...385...21K, 2014AJ....147...13K, 2015ApJ...802L..25T, 2015AJ....149..171T}.  The group has approximately 100 galaxy members, of which approximately 50\,percent have accurate distance estimates, mainly based on the HST program conducted by \citet{2007AJ....133..504K}, but also from various ground-based measurements  \citep{2000AJ....119..166J,2014ApJ...795L..35C,2016ApJ...823...19C}. The halo of the Centaurus Group extends over 3.5\,Mpc, from 3\,Mpc to 6.5\,Mpc along our line-of-sight (see Fig. 11 of \citealp{2017A&A...597A...7M}). Behind the group is a vast, empty region devoid of any matter called the Local Void {\citep{1987ang..book.....T}}. Because the Cen\,A subgroup is the most massive, gravitationally bound galaxy aggregate in the Local Volume (LV, \citealp{2004AJ....127.2031K,2013AJ....145..101K}) and  the M\,83 subgroup is less abundant and further away, it is intrinsically difficult to allocate dwarf galaxies unambiguously to a subgroup without accurate distance measurements. 

In \citet{2015A&A...583A..79M} we reported the discovery of 16 new dwarf galaxy candidates around M\,83 using wide-field imaging data collected with the Dark Energy Camera. This survey was subsequently extended to the entire Centaurus Group, leading to the detection of another 41 candidates \citep{2017A&A...597A...7M}.

An intriguing feature in the dwarf galaxy distribution of the Cen\,A subgroup has been recently reported by \citet{2015ApJ...802L..25T}: all but one dwarf galaxy around the host galaxy Cen\,A (= NGC\,5128) known at that time were aligned in two highly flattened planes (vertical scale height $rms\approx60$\,kpc). However, with the addition of the new dwarf galaxies only one of the two planes seems to be statistically significant \citep{2016A&A...595A.119M}. This plane of satellites shows some remarkable similarities to the planes of galaxies in the Local Group \citep[LG,][]{2012MNRAS.423.1109P,2013MNRAS.435.1928P,2013Natur.493...62I}  -- it is perpendicular to the prominent dust lane of Cen\,A; it contains an aligned stellar stream from the disrupted dwarf galaxy dw3 \citep{2016ApJ...823...19C}, and the extension of the plane is aligned with M\,83. 
However, the most striking feature is the correlation in phase-space.
Using the available line-of-sight velocities measured for half of the confirmed Cen\,A satellites, \citet{2018Sci...359..534M} discovered that most satellites with measured velocities share a coherent motion within the satellite plane, making it the third case of a co-orbiting satellite system in the local universe and the first outside of the LG. A quantitative comparison to cosmological simulations showed that this feature is as unlikely (<1\%) as it is in the LG. Hence the Centaurus Group of galaxies is an ideal environment for carrying out cosmological tests outside of the LG.

A fundamental observational ingredient to facilitate such cosmological predictions are accurate distances to dwarf galaxies. At a mean distance of $\sim 3.8$~Mpc \citep{2004AJ....127.2031K,2013AJ....145..101K} for the Cen\,A subgroup it is possible to resolve the brightest red giant stars with an 8m class telescope at a good site {\citep[e.g.\ ][]{2001A&A...379..781R,2006A&A...448..983R,2013MNRAS.432..832C}}. 
In this paper we present deep $VI$ stellar photometry based on the observations taken with the FORS2 instrument mounted at VLT for the three of our dwarf galaxy candidates (dw1325-33, dw1335-29, and dw1340-30), and establish tip of the red giant branch (TRGB) distances for two (dw1335-29 and dw1340-30). The third dwarf candidate dw1325-33 could not be resolved into individual stars.

\section{Observations and data reduction}
\label{obsred}

\begin{figure}[ht]
\centering
\fbox{\includegraphics[width=4.2cm]{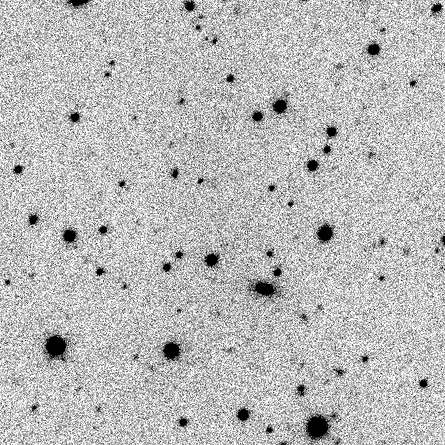}
\includegraphics[width=4.2cm]{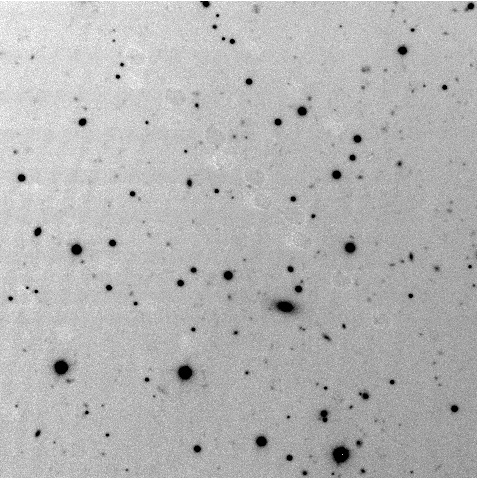}}\\
\fbox{\includegraphics[width=4.2cm]{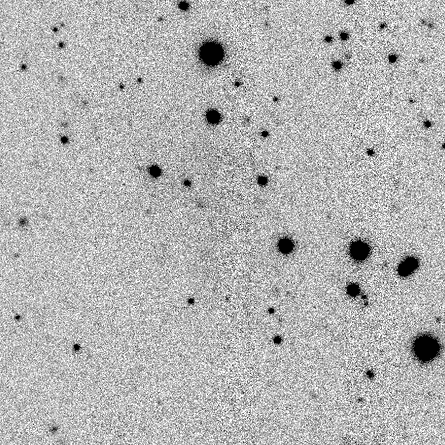}
\includegraphics[width=4.2cm]{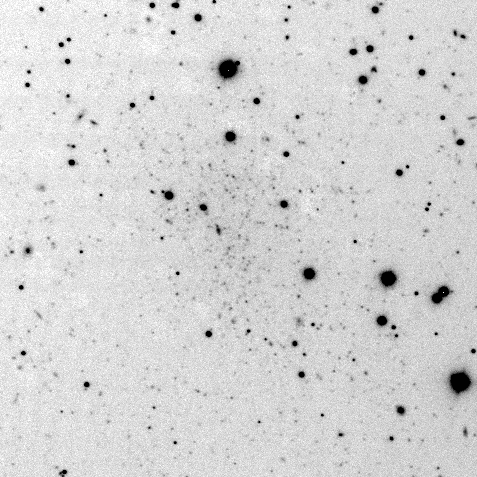}}\\
\fbox{\includegraphics[width=4.2cm]{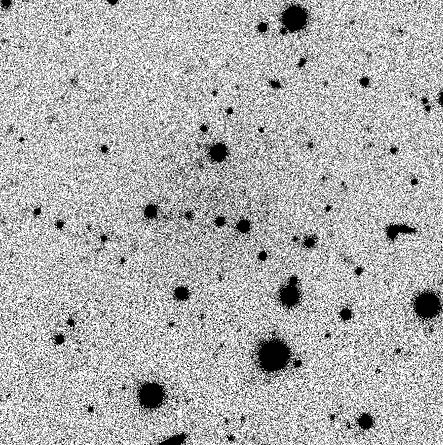}
\includegraphics[width=4.2cm]{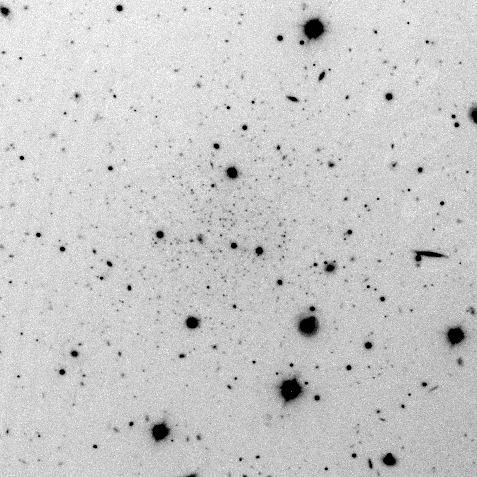}}\\
\caption{Left column: $r$-band DECam images of dwarf galaxy candidates dw1325-33 (top), dw1335-29 (middle) and dw1340-30 (bottom) taken from \citet{2015A&A...583A..79M}. Right column: deep, high resolution FORS2\@ VLT images of the same objects resolve the brightest stars in the dwarf galaxies, except in the case of dw1325-33 (see section \ref{casedw1325} for more details). All images are oriented with North up and East to the left. They have 2\,arcmin on each side.}
\label{taretsDECamVLT}
\end{figure}

Our targets were the three dwarf galaxy candidates dw1325-33, dw1335-29, and dw1340-30, a subsample from the 16 candidates found by our team in the M\,83 subgroup \citep{2015A&A...583A..79M}. In the left column of Fig.\,\ref{taretsDECamVLT} we show the discovery images observed with the wide-field Dark Energy Camera mounted at the prime focus of the Blanco 4-m telescope at CTIO.
A total of 5.2 hours of VLT time in Service Mode was subsequently allocated to the follow-up observing run 097.B-0306(A) under excellent observing conditions: dark time and seeing better than $0\farcs 6$, {which are necessary}  to resolve asymptotic giant branch (AGB) and {upper }red giant branch (RGB) stars at a distance of $\sim 5$~Mpc. Deep $V$ and $I$ CCD images were taken in 2016/17 with the FOcal Reducer and low dispersion Spectrograph (FORS2) mounted on the UT1 of the Very Large Telescope (VLT) of the European Southern Observatory (ESO). The FORS2 camera has a $6\farcm 8\times6\farcm 8$ field of view and is equipped with a mosaic of two $2k \times 4k$ MIT CCDs, which when used with standard resolution (SR) and $2\times 2$ binning offers a scale of $0\farcs 25$ per pixel. The targets were centered on Chip\,1, so that Chip\,2 is free from the stars of the galaxies and can be used as reference frame for estimating the contamination of Galactic foreground stars. 

In Table\,\ref{fit} we provide a log book of the science observations. The original request was for three $V$-band frames of 750\,sec and 12 exposures in $I$-band of 231\,sec for each target. A first set of $V$-band images for dw1335-29 was taken under seeing conditions slightly outside the requested constraint of $\lesssim 0\farcs 6$, and thus was repeated later. In the case of dw1340-30, a single $V$ band exposure was acquired with $0\farcs 7$ seeing before the sequence was aborted. The whole sequence was repeated later. These two sets of additional $V$ band images added additional frames and exposure time, which turned out to be useful for the stellar photometry. The combined best seeing {FORS2} images produced are shown on the right hand side of Fig.\,\ref{taretsDECamVLT}. 

\begin{table*}[ht]
\caption{Observation summary for the three dwarf galaxy candidates.}
\centering                          
\begin{tabular}{l l l l r l l l}        
\hline\hline                 
$\alpha_{2000}$ & $\delta_{2000}$ & Observing & Instrument & Exposure & Filter & Airmass & Image quality  \\    
(hh:mm:ss) & (dd:mm:ss) & Date &  & time (s) & & & (arcsec)\\    
(1) & (2) & (3)  & (4)  & (5) & (6) & (7) & (8)\\ 
\hline      \\[-2mm]                  
\bf{dw1325-33}\\
13:25:41 & $-$33:00:25 & 8/9 Apr 2016 & FORS2 & $3\times750 $ & V & 1.01 & 0.6\\
 &  & 8/9 Apr 2016 & FORS2 & $12\times231 $ & I & 1.12 & 0.6\\
\bf{dw1335-29}\\
13:35:46 & $-$29:43:50 & 7/8 Apr 2016 & FORS2 & $3\times750 $ & V & 1.16& 0.7\\
& & 7/8 Apr 2016 & FORS2 & $12\times231 $ & I & 1.07 & 0.5\\
 & 	 & 8/9 Apr 2016 & FORS2 & $3\times750 $ & V & 1.02& 0.5\\
\bf{dw1340-30}\\
13:40:19 &	$-$30:21:35 &  13/14 Apr 2016  & FORS2 & $12\times231 $ & I & 1.05 & 0.3 - 0.5\\
& &  29/30 Apr 2016  & FORS2 & $1\times750 $ & V & 1.01 &  0.7\\
& &  29/30 Jan 2017 & FORS2 & $3\times750 $ & V & 1.27 & 0.6\\
\hline
\end{tabular}
\tablefoot{(1)+(2): Coordinates of the candidates in the epoch J2000; (3) Date of observation; (4): Instrument used for the observation; (5): Exposures taken during the night; (6) Filter used for the observation; (7) Mean airmass during the observation; and (8): Average {image quality} measured on images.}
\label{fit} 
\end{table*}

\subsection{Data Reduction and Calibration}
\label{calibration}
Data reduction was carried out within the Image Reduction and Analysis Facility (IRAF) software \citep{1993ASPC...52..173T}. A median combined stack of 10 bias frames was used to create a master bias. This was subtracted from all other images. Master flat fields were then created by combining five twilight flats for each filter using an average sigma clipping algorithm to reject any residual faint stars, bad pixels and cosmic rays. These {calibration} images where then used to process individual science frames and the photometric standard star fields.

The photometric calibration was based on standard star observations from the ESO calibration plan. The photometric standard star fields taken in the $VI$-bands during each of our observing nights are listed in Table~\ref{tab:standards}\footnote{Standard star observations from 29/30 Jan 2017 were not used. We preferred to calibrate both $V$ and $I$-band images of dw1340-30 with respect to the April 2016 observations, due to the lack of $I$-band observations after the M1  mirror recoating in August 2016, which would introduce an unknown color-term.}. We derived the zero points $ZP_{V,I}$ and extinction coefficients $k_{V,I}$ (Table\,\ref{tab:zp}) by comparing our standard star measurements to the values in the Stetson standard star catalogue \citep{2000PASP..112..925S}, and found them to be consistent with values available from the ESO Quality Control webpages, as well as with an independent cross-calibration based on our DECam images. The photometry was calibrated using the formulae:
\begin{eqnarray*}
V &=& V_{instr} + ZP_V - k_V\cdot X_V - A_V + 2.5\cdot \log_{10}(t)\\
I &=& I_{instr} + ZP_I - k_I\cdot X_I - A_I + 2.5\cdot \log_{10}(t)
\end{eqnarray*}

\noindent where the exposure time $t$ is given for a single (reference) exposure (750 and 231 seconds for $V$ and $I$, respectively), $A_V$ and $A_I$ are the Galactic extinction values based on the reddening map by \citet{1998ApJ...500..525S} and the correction coefficients from \citet{2011ApJ...737..103S},  and the airmass ($X_V,X_I$) of a chosen reference exposure is used.

\begin{table}[ht]
\caption{Observation log for standard star fields.}
\centering                          
\begin{tabular}{l l l l}        
\hline\hline                 
Date & Field & Airmass & Filters\\
\hline
7/8 Apr & PG 1323 & 1.07 & V \& I\\
 2016   & IC 4499 & 1.87 & V \& I\\
\hline
       & NGC 2437 & 1.02 & V \& I\\ 
8/9 Apr& E5       & 1.62 & V \& I\\
 2016  & E7       & 1.27 & V \& I\\
       & E7       & 1.08 & V \& I\\
\hline
13/14 Apr& NGC 5139 & 1.10 & I \\
 2016    & IC 4499  & 1.87  & I \\
\hline
 29/30 Apr& NGC 2818 & 1.03  & V \\
2016    & NGC 5139 & 1.78  & V \\
          & E7  & 1.26  & V \\              
\hline
\end{tabular}
\label{tab:standards}
\end{table}

\begin{table}[ht]
\caption{Calibration coefficients.}
\centering                          
\begin{tabular}{l r r r}        
\hline\hline                 
 & dw1325-33 & dw1335-29 & dw1340-30\\
\hline
$ZP_V$ (mag)& 27.889 (0.003) & 27.887 (0.003) & 27.888 (0.004)\\   
$ZP_I$ (mag)& 27.334 (0.004)& 27.343 (0.004)& 27.333 (0.003) \\     
$k_V$ & 0.121 (0.003)& 0.108 (0.001) & 0.132 (0.003)\\     
$k_I$ & 0.056 (0.003) & 0.068 (0.002)&  0.075 (0.002)\\    
\\
$X_V$ & 1.011 &1.010 & 1.050\\
$X_I$ & 1.157 &1.093& 1.065\\
$A_V$ (mag)& 0.147& 0.125& 0.158\\
$A_I$ (mag)& 0.081& 0.069& 0.087\\
\hline
\end{tabular}
\label{tab:zp}
\end{table}

\subsection{Photometry}
We carried out photometric measurements for all stars on the bias subtracted and flat-fielded science images using the standalone version of DAOPHOT2 \citep{1987PASP...99..191S} package, which is particularly well suited for crowded fields. 
The steps carried out include detection of point sources, aperture photometry with a small aperture and then point-spread function (PSF) modeling. We modeled a PSF using 50 isolated bona fide stars across the field on every individual science frame. To check the quality of  the PSF model we visually inspected the residual images of all chosen stars and rejected those with a strong residual. From the best $V$ and $I$ frames ($fwhm < 2.3$\,px on average for unsaturated stars) a single deep image was produced using MONTAGE2. On this deep image the DAOPHOT2 routines FIND, PHOT and ALLSTAR were run to produce the deepest possible point source catalog. This catalog was then used as input for simultaneous PSF fitting on every science frame  using ALLFRAME. The resulting catalogs for each individual image were average combined per filter with the DAOMATCH and DAOMASTER routines keeping only those measurements that had been detected on 2/3 of all input images for the given filter. Finally, stars detected within 1~pix tolerance from both $V$ and $I$ lists were combined to make the master catalogue. 

\begin{figure}[ht]
\centering
\includegraphics[width=9.9cm]{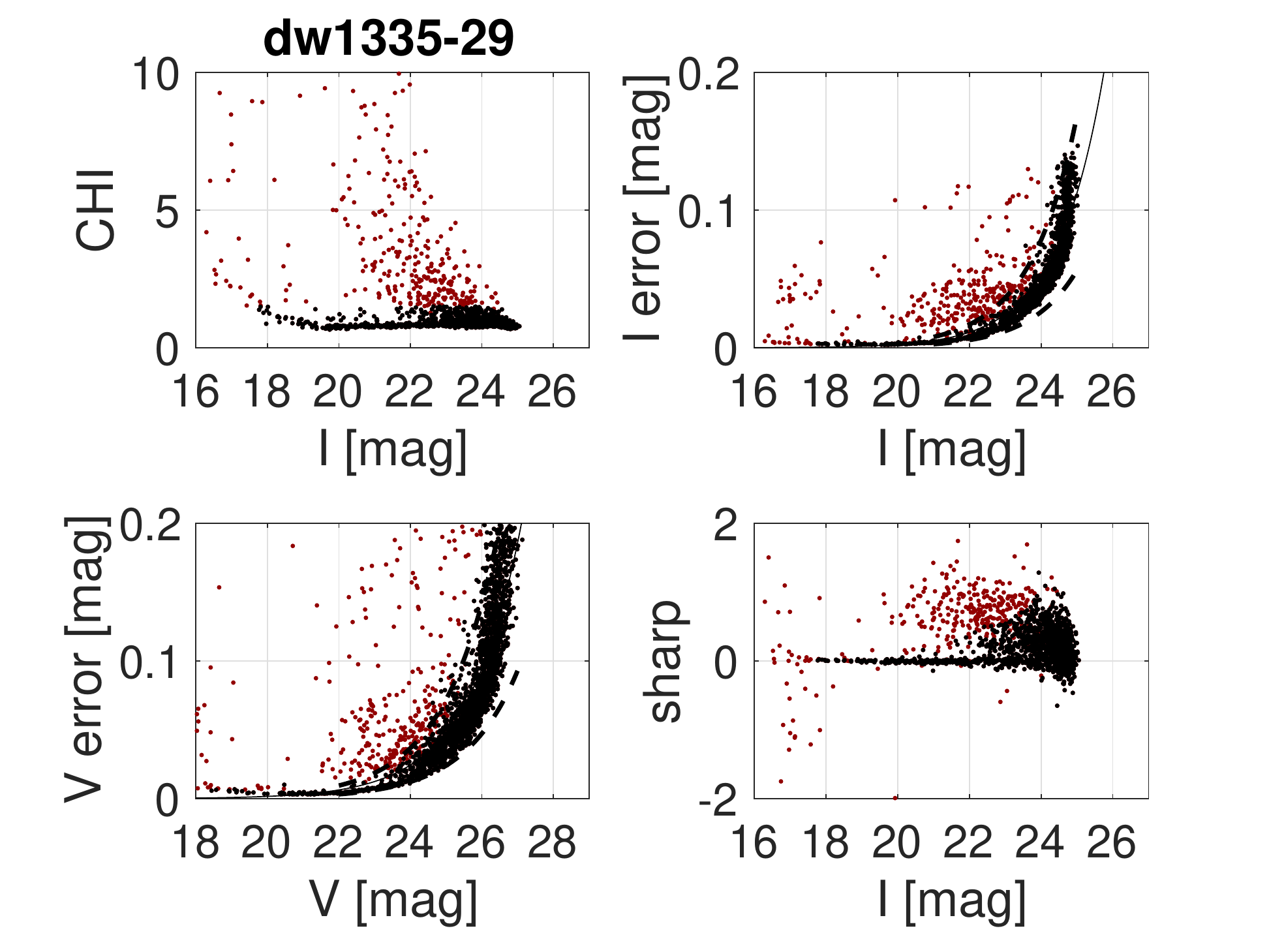}\\
\caption{Selection of bona fide stars from the master catalogue of dw1335-29 (see text). The black dots correspond to stars that fulfill all point source constraints and remain in the catalog, while the red dots are rejected objects.}
\label{goodstars}
\end{figure}

To further purge the master catalog from non-stellar or blended sources, only stars with good shape parameters were included. The constraints were the following: the value for $\chi$ had to be smaller than 1.5, $sharpness$ had to be between $-2$ and $2$, the magnitude errors were only allowed to deviate 50\% from the best fitting value at a given magnitude. {To explain the latter more precisely: we fitted an exponential function through the magnitude errors (as function of the measured magnitude) and defined the intervals from 0.5 to 1.5 times the fitted value as the locus {within which we keep the detected objects as bona-fide stars}. In total a}  star has to fulfill all {these} constraints to remain in the master catalog. See Fig.\,\ref{goodstars} for an illustration of the different constraints in the case of dw1335-29. The plume of objects with high $\chi$ values (red symbols in the upper left panel of Fig.\,\ref{goodstars}) and positive $sharpness$  values are likely background galaxies and thus are removed from the catalog.

\subsection{Completeness and error analysis}
\label{artStarErr}
\begin{figure}[ht]
\includegraphics[width=9cm]{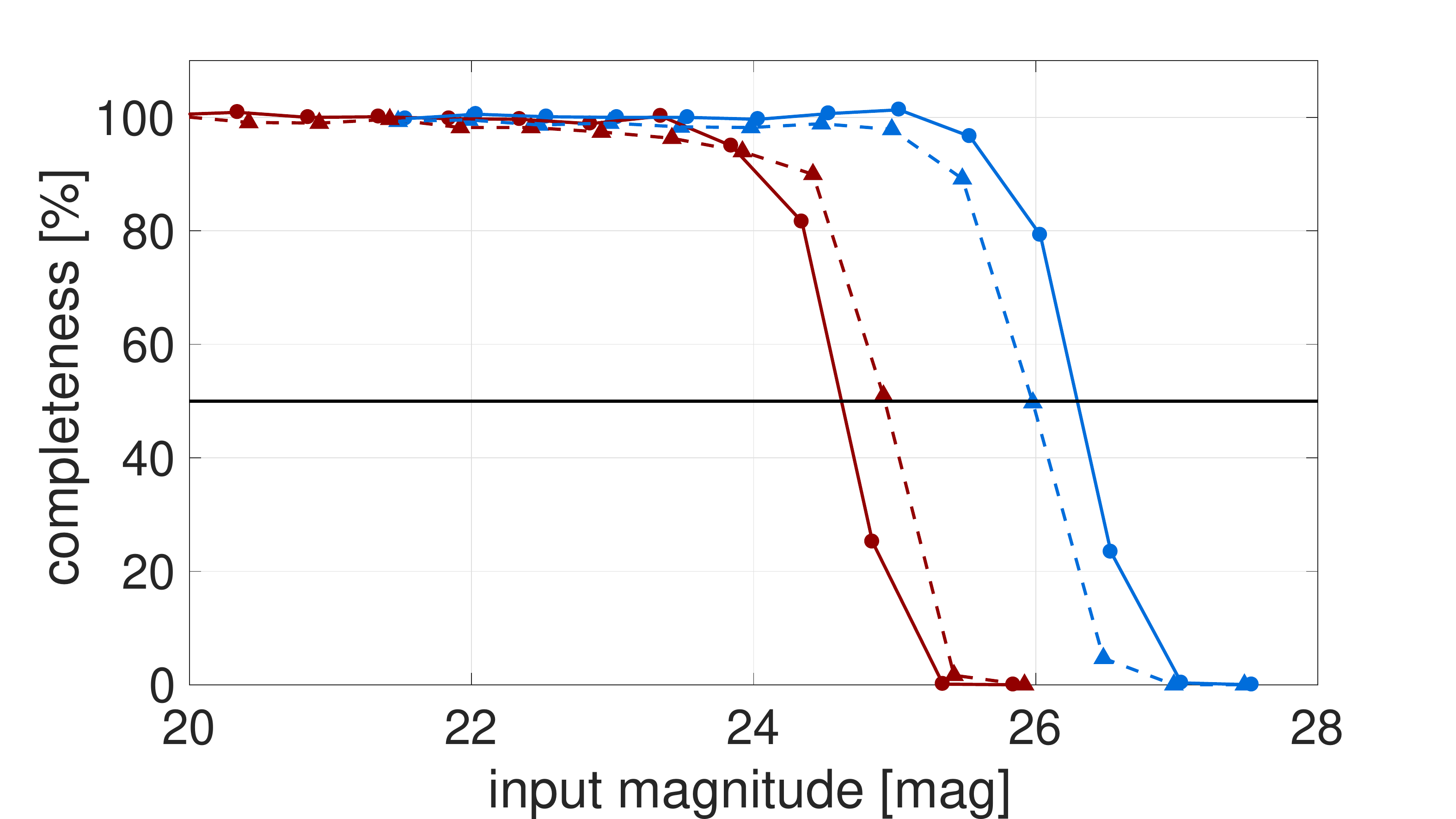}
\includegraphics[width=9cm]{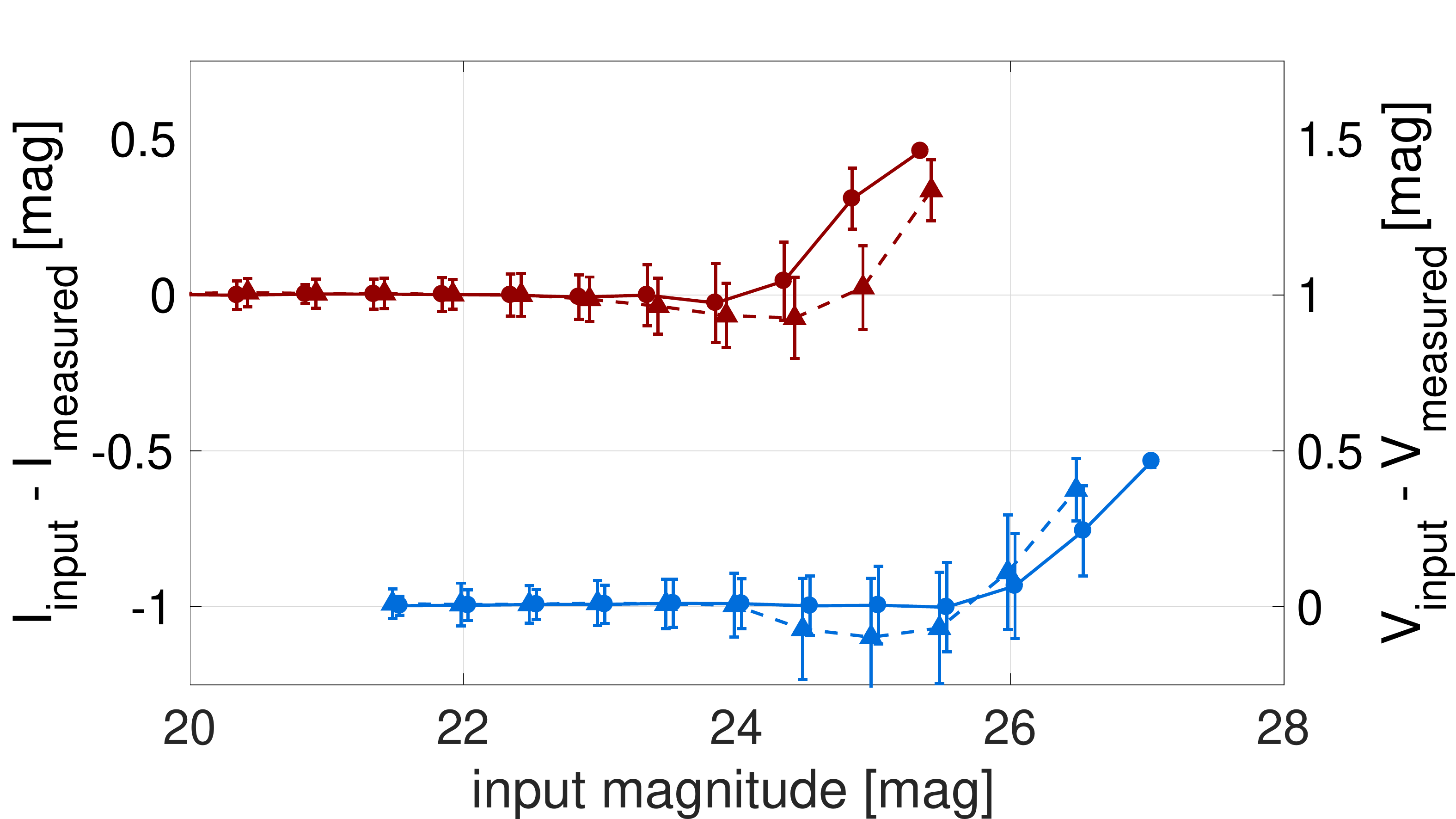}
\caption{{Results of our artificial star tests for $I$ (red) and $V$ (blue) bands (circles and straight line for dw1335-29, triangles and dashed line for dw1340-30).} Upper:  The recovery rate of the artificial stars induced into the science frames as a function of the input magnitude. The 50\% level is highlighted with the black horizontal line. Bottom: The difference between the input magnitude and measured magnitude as a function of input magnitude.}
\label{artTests}
\end{figure}

We performed artificial star tests to assess the detection completeness and characterize the measurement errors for our PSF photometry. For that purpose, we generated 900 artificial stars, all with the same magnitude for each 0.5\, magnitude bin between 20--28\,mag. These stars were then uniformly spread over the field and added to every individual science frame, distributing them on a hexagonal grid with a random starting position. The angular separation between stars corresponds to twice the PSF fitting radius, such that the artificial stars do not increase the natural crowding. The dithering offsets between frames were carefully taken into account to ensure that each artificial star is added in every science frame at the exact same sky position. The same PSF photometry pipeline as used before is then applied on every {image with artificial stars added for each} magnitude iteration of the interval, with the only exception that the PSF model constructed in the scientific photometry is used as there is no need to re-derive it. The recovery rate for the artificially created stars, together with the difference between input and output magnitudes are presented for dw1335-29 and dw1340-30 in Fig.\,\ref{artTests}. The plot on the top presents the completeness curves of our star detection algorithm, the bottom plot gives the error of our PSF photometry as a function of input magnitude. 

For the dw1335-29 data the completeness drops below 50\% at 24.6\,mag  in $I$ and at 26.3\,mag in $V$. The measured PSF-based photometry error starts to grow at $24.0$\,mag for $I$, and $25.5$\,mag for $V$. 
For dw1340-30 the completeness drops below 50\% at 24.9\,mag in $I$ and 25.9\,mag in $V$, the error starts at 24.3\,mag in $I$ and 24.9\,mag in $V$, respectively. {We note that for dw1340-30 the error drops before the final rise, meaning that we underestimate the measured magnitude. However, within the uncertainty the mean value still agrees with unity. At these magnitude steps we find a systematic dependence of the photometry along the y-axis of the chip which is not apparent within other magnitude steps. We further note that we reach a deeper $I$ band limit for dw1340-30, which is due to better seeing conditions.}

\begin{figure*}[ht]
\centering
\includegraphics[width=9cm]{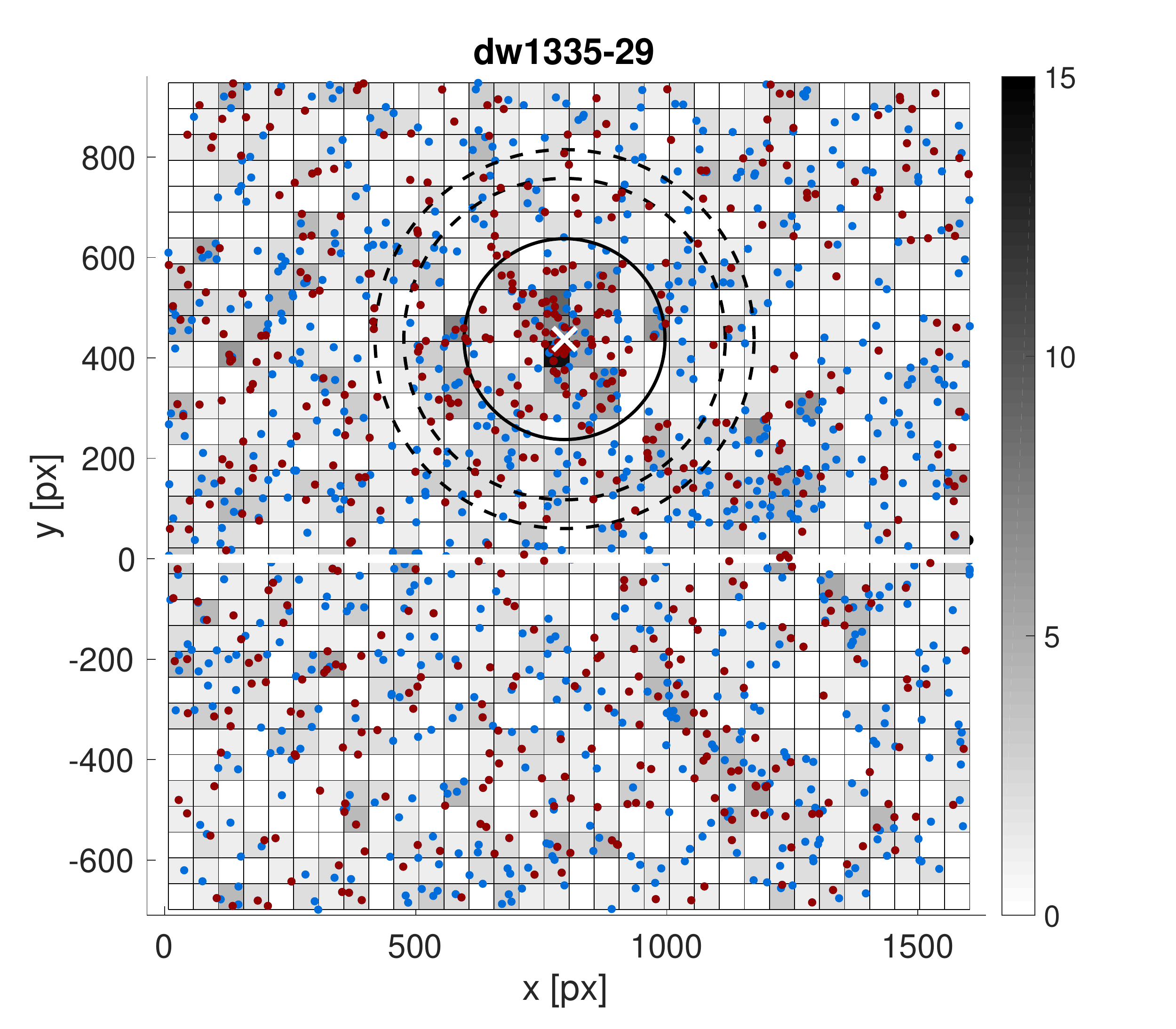}
\includegraphics[width=9cm]{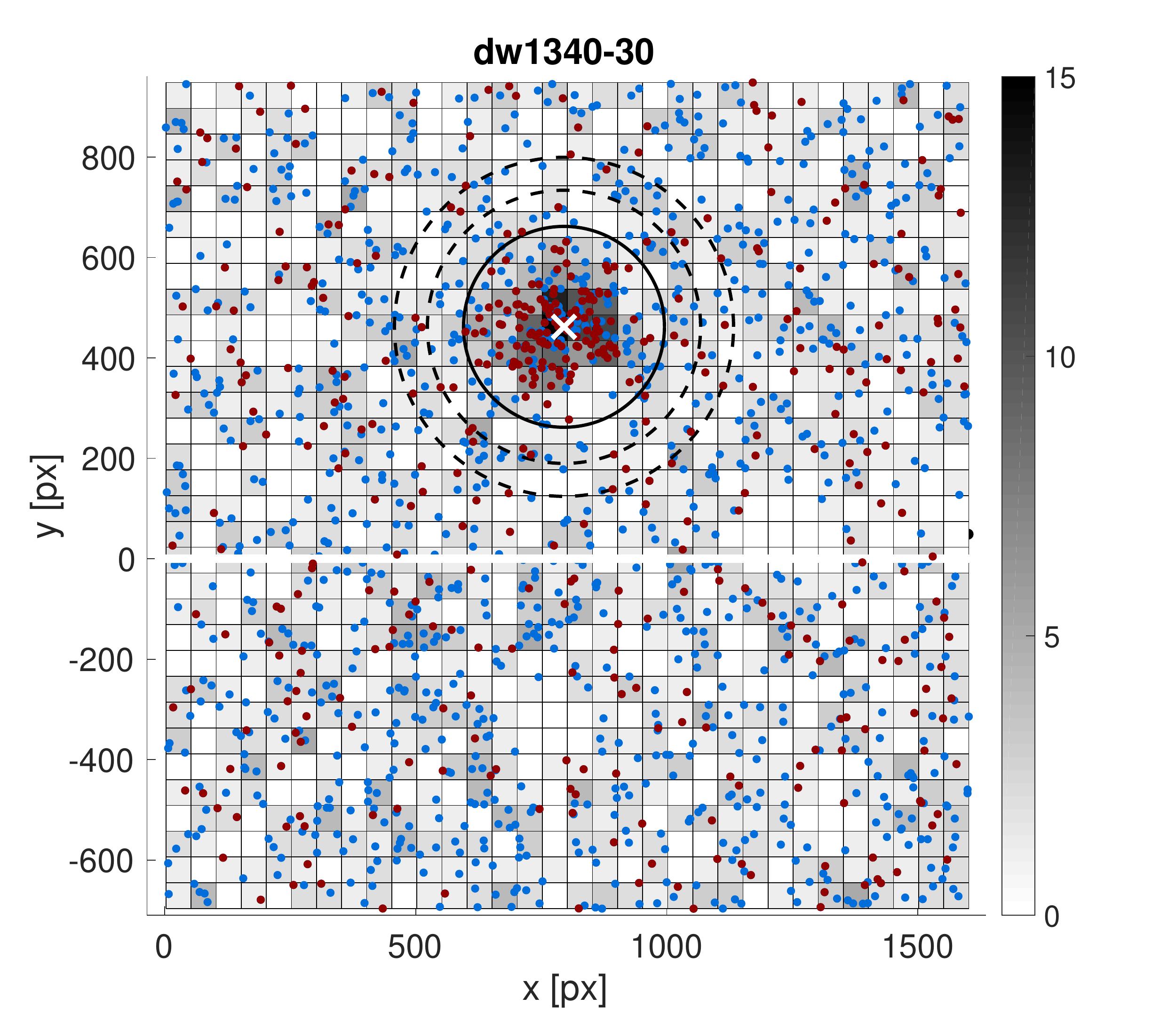}
\caption{Distribution of stars (red and blue dots) with $I$-band magnitudes in the range 24 to 25.5, underlaid by a 2D histogram that represents the local star density (stars per 156 arcsec$^2$). The cross marks the nominal center of the galaxy and the circle has a radius of $50\arcsec$ (=200\,px). The left panels of Fig.\,5 shows the CMD of all stars detected within that radius. The two outer circles (dashed lines) define an annulus with the same area as the inner circle. This area is used for statistical decontamination of the CMDs, and the annulus has the inner radius at the location corresponding to four times the effective radius \citep{2015A&A...583A..79M} of the corresponding galaxy. Stars in the annulus were used to construct the CMD of Galactic foreground stars in the direction of the dwarf galaxy.}
\label{colormap}
\end{figure*}

\subsection{The case of dw1325-33}
\label{casedw1325}
\begin{figure}[ht]
\includegraphics[width=9cm]{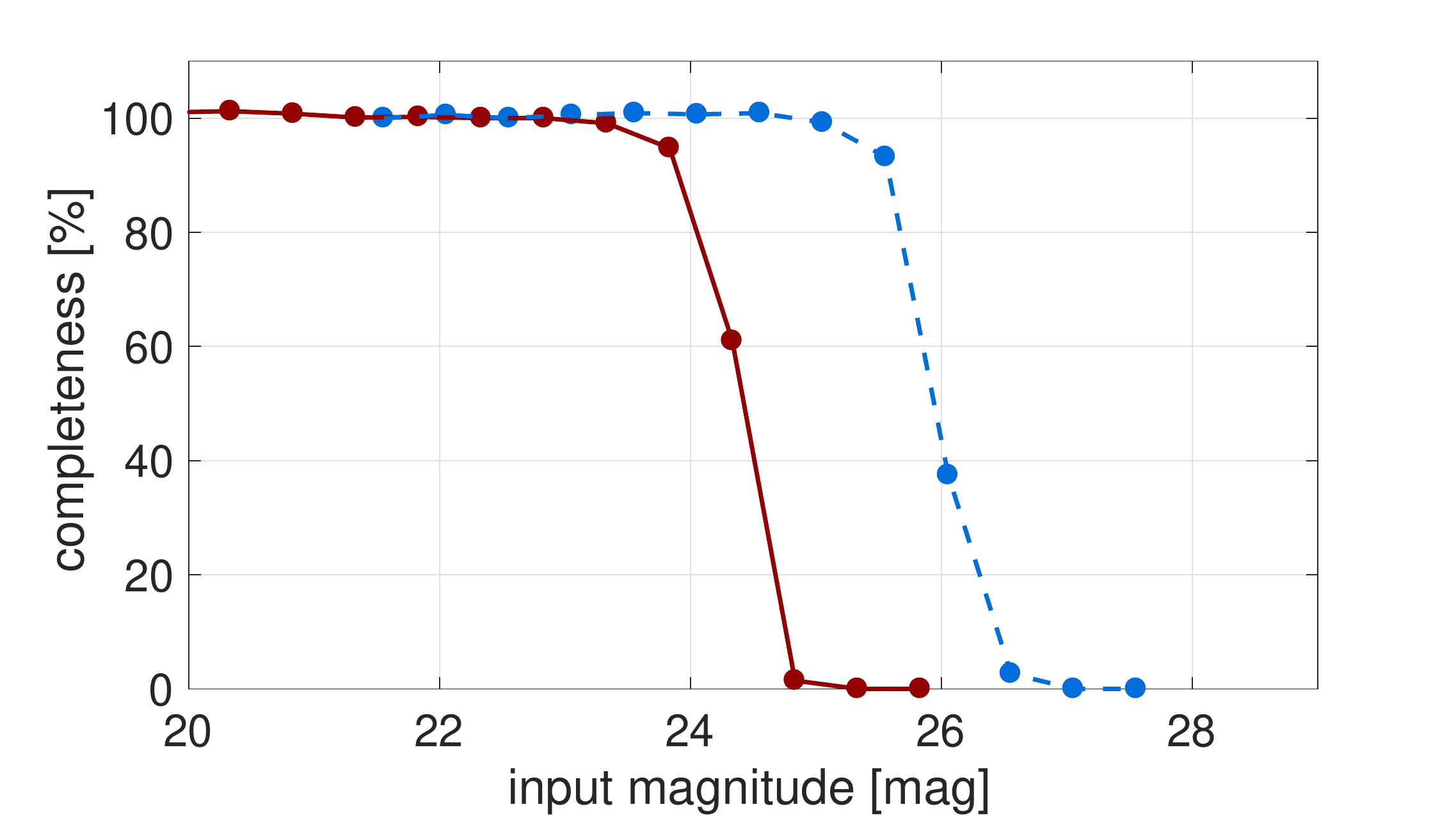}
\caption{Results of our artificial star tests for $I$ (red line) and $V$ (blue dashed line) bands for dw1325-33.}
\label{artTestsdw1325}
\end{figure}

{The third dwarf galaxy candidate, dw1325-33, could not be resolved into {individual} stars with our data. The reason is not as simple as the candidate not being a nearby dwarf galaxy{, although this is also a possible explanation}. {The presence of} a higher underlying sky background (in excess of 16,700 ADU compared to $\sim 13,000$ ADU in the other two galaxies) prevented detection of faint stars in the images. There is also a strong radial gradient in the background light. We modeled this with IRAF's imsurfit command and stacked the resulting images. There is indeed a low-surface brightness feature at the position of the candidate, but the quality of the image is too {low} to perform {reliable photometric measurements}. With our current observations we {cannot rule out nor confirm} the possibility that this candidate is a member of the Centaurus group.} We applied our artificial stars analysis to estimate the detection limit within the field of dw1325-33, see Fig.\,\ref{artTestsdw1325} for the results. We are not able to detect stars (10\% detection rate) fainter than 24.8\,mag, corresponding to a distance of 6{.0}\,Mpc for the TRGB. Considering the steeply dropping completeness below the 50\% limit of I=24.4 \,mag, we conservatively set a lower limit for the dw1325-33 distance to 5.3\,Mpc.

\section{Results}
In the following subsections we re-detect the dwarf galaxies as stellar over-densities in the deep CCD images, present their color-magnitude diagrams and measure the TRGB distance from the $I$-band luminosity function.

\subsection{Stellar overdensity}
\label{overdensity}
When star clusters and nearby dwarf galaxies are resolved into individual stars they can be detected as stellar overdensities  superimposed on the Milky Way's foreground star population \citep[e.g.][]{2007ApJ...654..897B,2015ApJ...799...73K,2016ApJ...820..119K}. Similarly, but given the much smaller spatial extent and thus significantly lower amount of foreground+background contamination, we blindly detect the dwarf galaxy candidates dw1335-29 and dw1340-30 in our CCD images as pronounced stellar overdensities.

In Fig.\,\ref{colormap} we show the distribution of all stars within $24<I<25.5$\,mag. In this magnitude range we expect the TRGB stars at the distance of the M\,83 subgroup (see below). The star density is shown in gray scale and sampled on a grid using $50\times50$\,px bins. There are clear overdensities of stars at the positions of the dwarf candidates, as expected if the stars are associated to the galaxies and resolved into individual stars. We note that we find the same overdensity when removing the magnitude constraint for the stars. The estimated radius of the galaxy, where the overdensity is not distinguishable from the foreground population is indicated as solid circle (radius of $50\arcsec$ corresponding to 200\,px), in comparison to the estimated half light radii of $20\arcsec$ and  $17\arcsec$ for dw1335-29 and dw1340-30, respectively \citep{2015A&A...583A..79M}.

To determine the center of the stellar overdensity we used a $k$-means algorithm, which samples a set of data into $k$ clusters and minimizes the sum of the distances of each data point to the center of the cluster. Using $k=1$ we can exploit this algorithms capability to find the center of the galaxy. The centers are at ($\alpha_{2000}$=13:35:46, $\delta_{2000}$=$-$29:42:28) for dw1335-29 and ($\alpha_{2000}$=13:40:19, $\delta_{2000}$=$-$30:21:31) for dw1340-30, respectively, which agree with the centers estimated in \citet{2015A&A...583A..79M} based on galaxy surface photometry. 

\subsection{Color-magnitude diagrams}
\begin{figure}[ht]
\includegraphics[width=9cm]{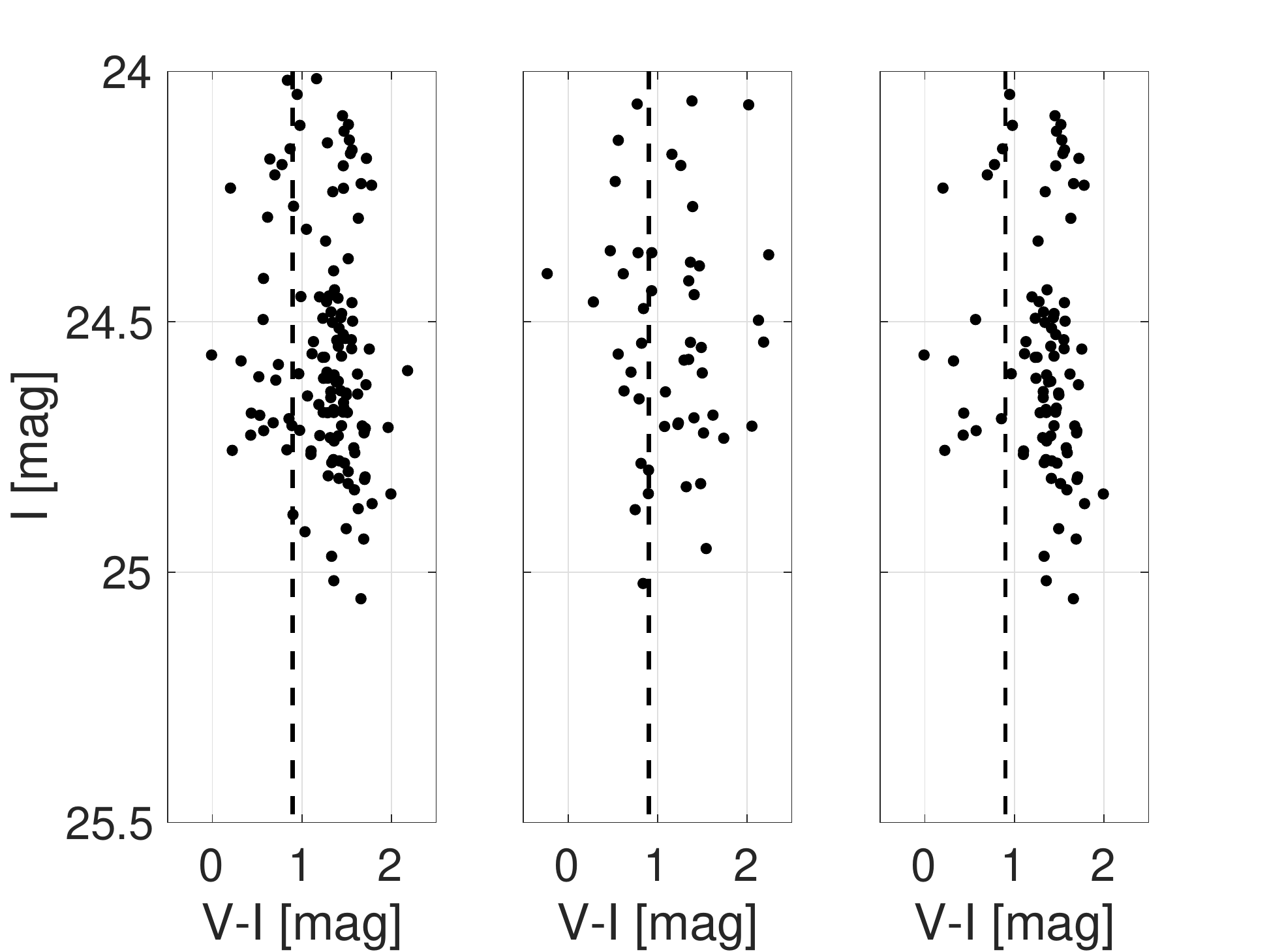}
\includegraphics[width=9cm]{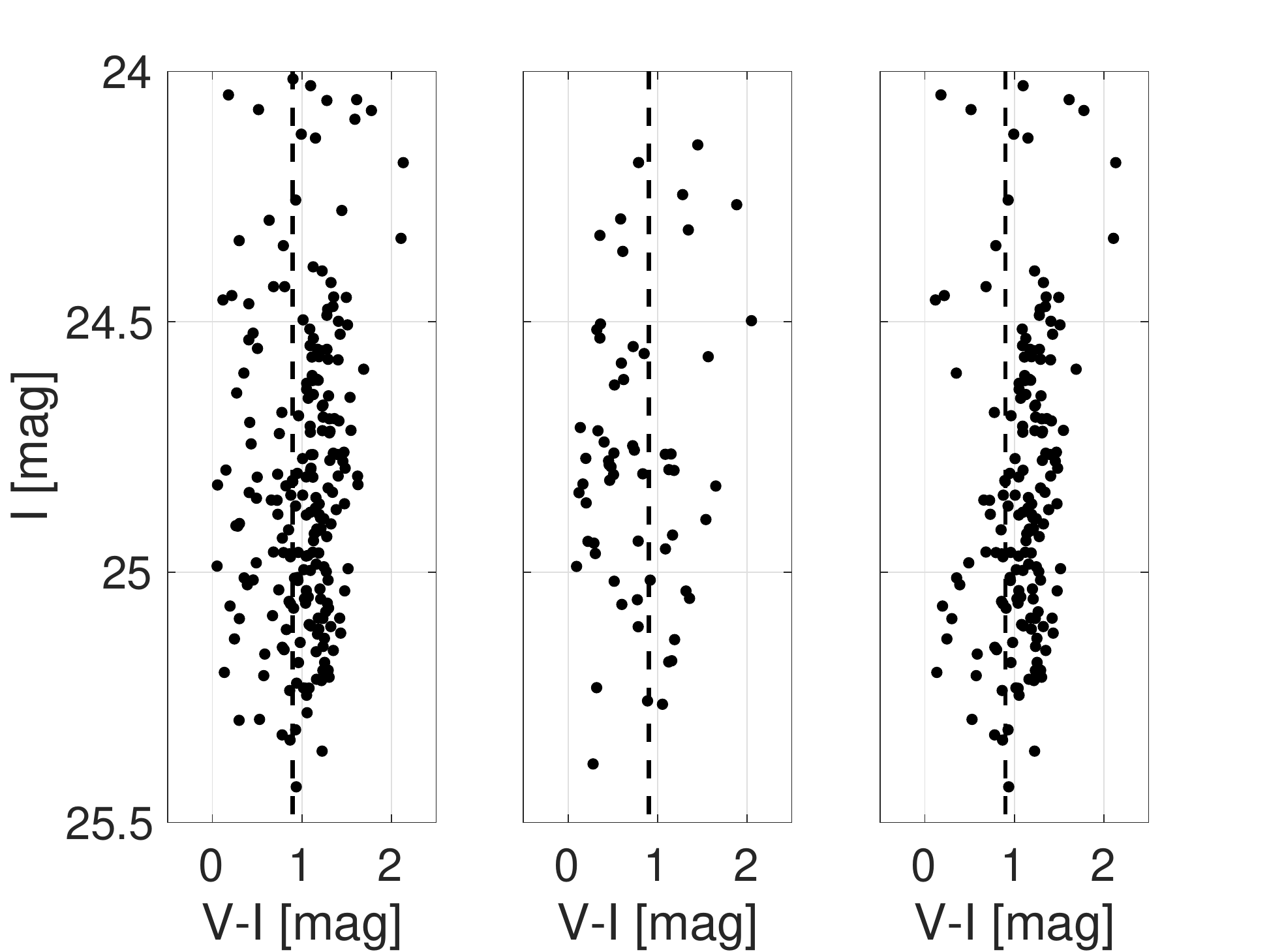}
\caption{Extinction-corrected color-magnitude diagrams of dw1335-29 {(top)} and dw1340-30 {(bottom)}. Indicated with dashed lines is the $V-I<0.9$\,mag boundary. Left: CMD for stars inside a radial aperture of {$50\arcsec$}. {Middle: Stars within the reference aperture used for the background subtraction. Right: Stars remaining after statistical cleaning of the CMD.}}
\label{cmds}
\end{figure}

Having performed the stellar photometry on Chips\,1 and 2 in Section\,\ref{obsred} and re-detected the galaxies as stellar over-densities in Section\,\ref{overdensity} it is possible to establish the color-magnitude diagram (CMD) for each dwarf galaxy. For that purpose, we considered all stars within a circle of radius $50\arcsec$ (indicated as circle in Fig.\,\ref{colormap}), corresponding to the approximate extent of the galaxy ($2.5 r_h$). 
Another CMD was created {for} the reference field {area, which has been selected to have the same size by} placing {an annulus around the galaxy (indicated as dashed ring in Fig.\,\ref{colormap})}, where the stellar population of the dwarf galaxies should not affect the CMD. Both CMDs are presented in Fig\,\ref{cmds}. Comparing the field CMD to the galaxy CMD it becomes clear that a significant number of foreground stars contaminates the CMD of the galaxy. 

To clean the CMD of the galaxy from foreground stars we statistically subtracted stars using the field CMD. For every star in the field CMD the star with the closest matching magnitude and color from the galaxy CMD was removed, but only if it the magnitude difference was smaller {than the photometric error at given magnitude (a superposition of the systematic error and its corresponding standard deviation, see Fig.\,\ref{artTests})}.  As the field stars are evenly distributed and we use the same area for both CMDs this should statistically clean the galaxy from foreground stars.  
In Fig.\,\ref{cmds} {right side} we present the remaining stars {after the subtraction}. In the case of dw1335-29 {90} stars remain in the CMD and {42} stars are removed, for dw1340-30 {167} stars remain and {51} stars are removed. The high number of subtracted stars is not surprising as the targets are relatively close to the Milky Way plane ($b=32.2$, and $b=31.3$). Additionally the projected distance to M\,83 is only $\sim$20\,kpc, and $\sim$85\,kpc for dw1335-29 and dw1340-30, respectively, meaning that there is some contribution of M\,83 stars expected in the CMDs. {This presence of M\,83 extended halo stars that may vary across the field is an additional reason to select the "field" area not too far from the dwarf. }

\subsection{TRGB distance}
\label{trgb}
The tip of the red giant branch (TRGB) method has become the method of choice to measure distances in the Local Volume {since the method was established in the early 90-ies} \citep{1990AJ....100..162D,1993ApJ...417..553L}. There is a sharp cut-off in the luminosity function at the bright end of the first ascent giant branch. This is a well-understood phenomenon: on their evolutionary path the red giants leave the red giant branch after the helium flash occurs. This physical process can be exploited as a standard candle. The turn-off of the red giant branch happens at the absolute magnitude of $M_I\sim -4$. The TRGB method is on par with RR Lyrae and Cepheid distance estimates with a typical uncertainty of 5\,percent {\citep{1993ApJ...417..553L,2009ApJ...697..996M,2016ApJ...832..210B}}.

\begin{figure}[H]
\includegraphics[width=9cm]{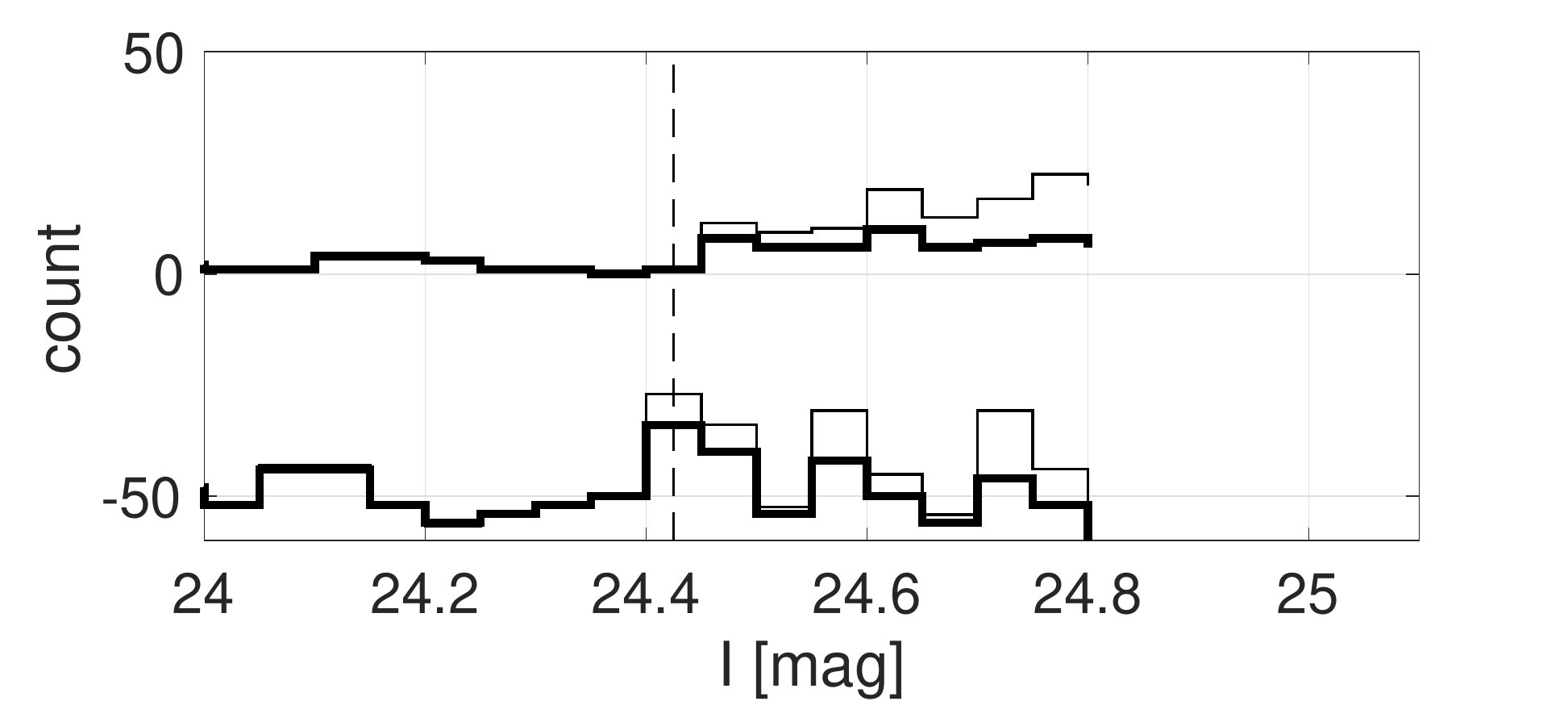}
\includegraphics[width=9cm]{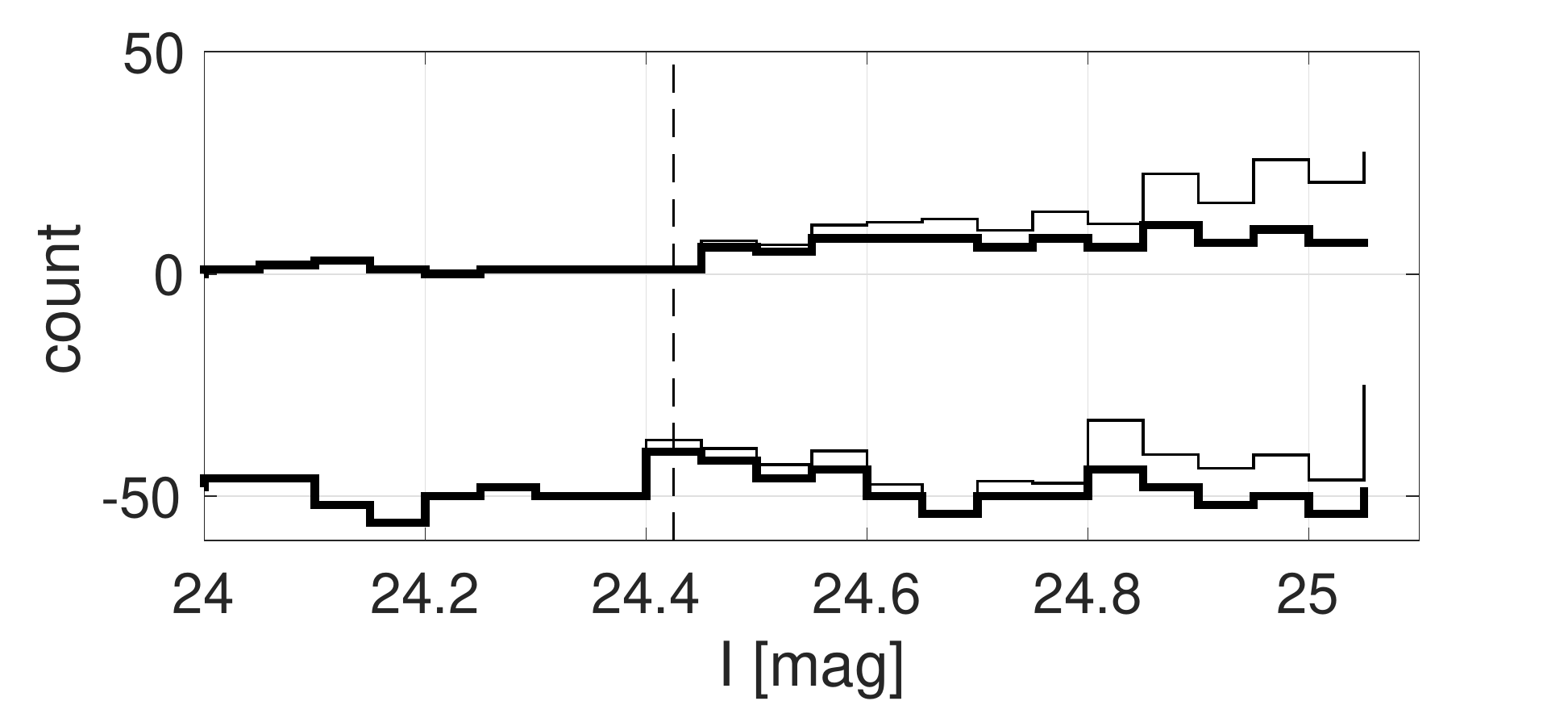}
\caption{The $I$-band stellar luminosity function for dw1335-29 (top) and dw1340-30 (bottom). Each panel shows the observed (bold line) and completeness-corrected 
(thin line) luminosity functions to the point where the completeness drops bellow 30\%.
The corresponding signals from the Sobel edge detection are shown below the respective luminosity function. The maximum (dashed line) indicates the location of the TRGB.}
\label{LF}
\end{figure}

In practice the tip magnitude of the RGB can be estimated by constructing the $I$-band luminosity function for all stars in the galaxy ($0.9<V-I<2.0$) and convolving it with a Sobel edge kernel $[-2,0,2]$ \citep{1993ApJ...417..553L}. The maximum of the convolution indicates the position of the TRGB, while the width of the peak is an estimate of the uncertainty. The LF was constructed using a bin width of 0.05\,mag, see {Fig.\,\ref{LF}}. {The {statistically} subtracted CMD was used to construct the LF.} The maximum of the edge detection is measured for both galaxies, dw1335-29 and dw1340-30 at $I=24.425$\,mag. {We use the middle of the bin to estimate the edge.}

\begin{figure}[ht]
\includegraphics[width=4.4cm]{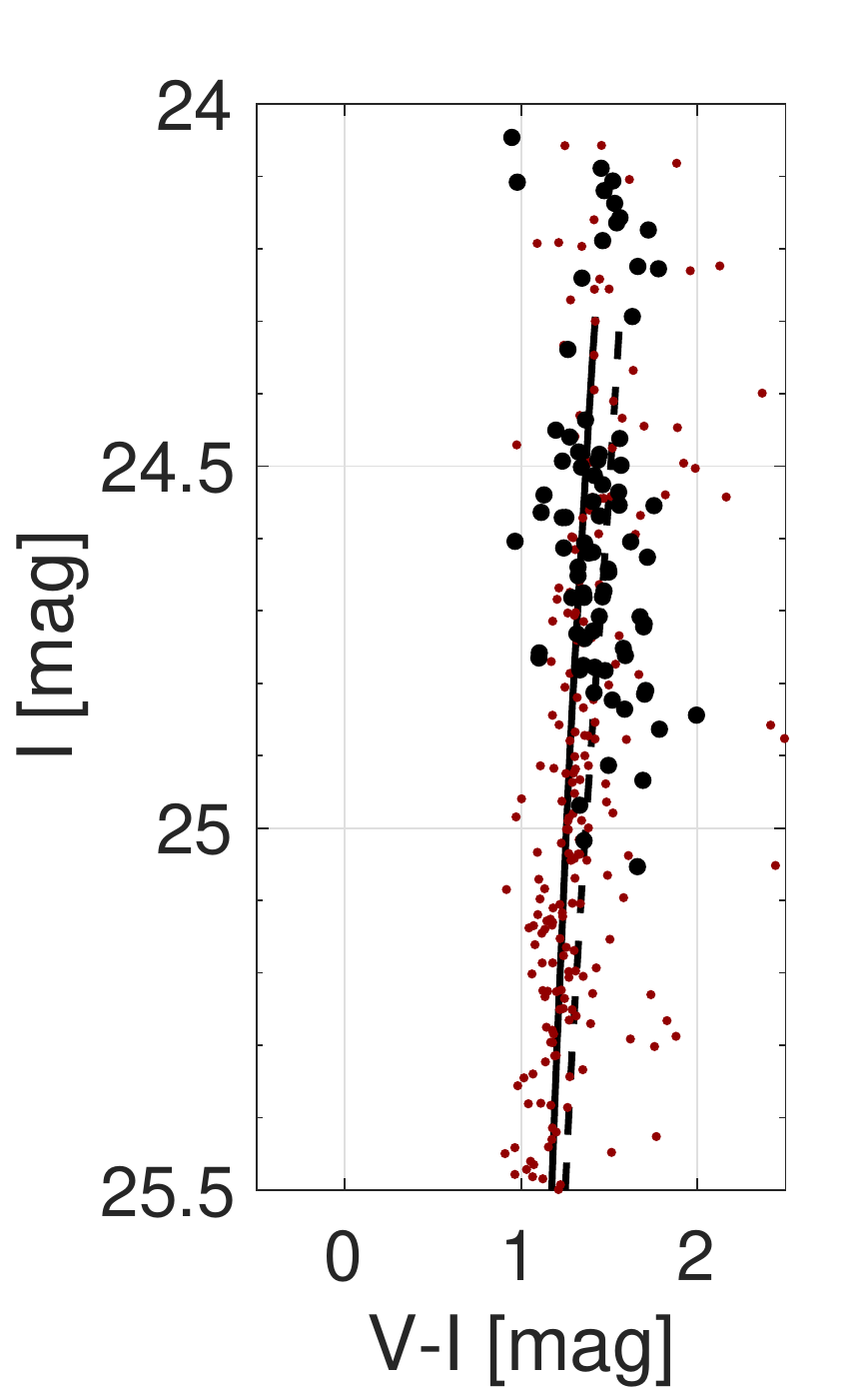}
\includegraphics[width=4.4cm]{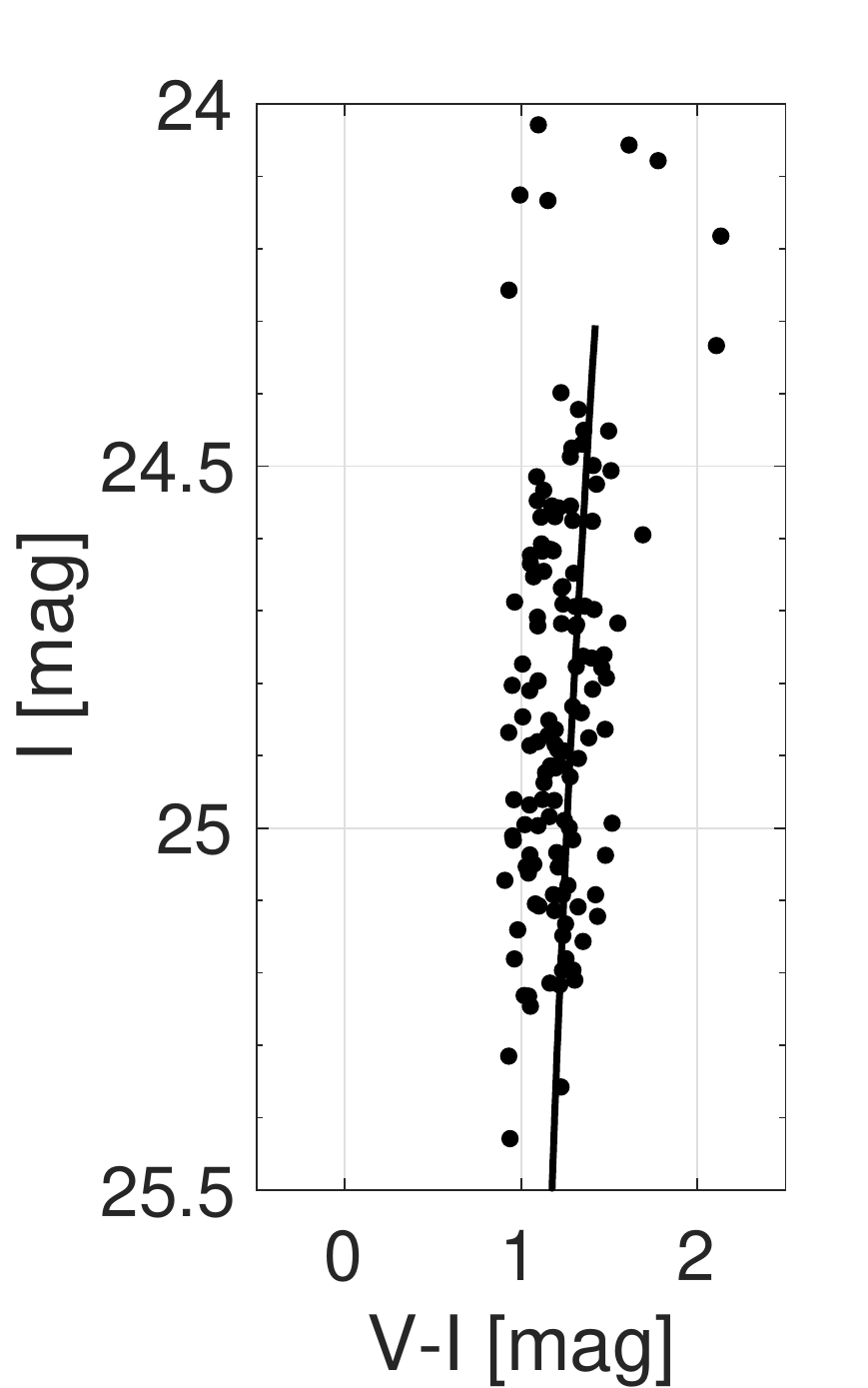}
\caption{Cleaned, extinction-corrected color-magnitude diagram with the best-fitting isochrone (line) {for dw1335-29 (left) ad dw1340-30 (right). For dw1335-29 we overplot the HST photometry (red) kindly provided by Andreia Carrillo. The dashed line corresponds to the best-fitting isochrone [Fe/H]=-1.266 by \citet{2017MNRAS.465.5026C}.}}
\label{isoBootstrap}
\end{figure}

To calibrate the derived TRGB magnitudes we used the equations provided by \citet{2004A&A...424..199B} and \citet{2007ApJ...661..815R}:
\begin{eqnarray*}
 M^{TRGB}_I&=&-4.05 +0.217[(V-I)_0-1.6]\\
 (V-I)_0&=&0.581 \rm{[Fe/H]}^2 + 2.472 \rm{[Fe/H]}+4.013
 \end{eqnarray*}
To estimate the mean metallicities [Fe/H] for the dwarfs we fitted theoretical BaSTI isochrones \citep{2004ApJ...612..168P} {to mean RGB color (Fig.\,\ref{isoBootstrap})}. 
For dw1335-29 a metallicity of $[$Fe/H$]= -1.79$\,dex is estimated, corresponding to a calibrated absolute TRGB magnitude of $M_{I}=-4.08\pm${0.02}\,mag, which gives a final distance of 5.03$\pm${0.24}\,Mpc. For dw1340-30 the metallicity is $[$Fe/H$]= -2.27$\,dex, the calibrated absolute TRGB magnitude of $M_{I}=-4.09\pm${0.02}\,mag and a final distance estimate of  5.06$\pm${0.24}\,Mpc. These distances put them within the virial radius of M\,83 \citep{2015A&A...583A..79M}, making them satellite galaxies of M\,83.
Information on the dwarfs are shown in Table\,\ref{properties}. {The estimated metallicities lie well within the metallicity - luminosity relation for dwarf galaxies \citep{2012AJ....144....4M}.}

\subsection{TRGB error estimation}
{The error budget for the TRGB magnitude comes from different sources, i.e. the uncertainties in the zero points, the PSF photometry, the TRGB calibration and the Sobel edge detection. The error in the zero point estimation derives from the standard frames uncertainties (Section.\,\ref{calibration}) and is $I_{ZP,error}=0.02$\,mag. The error in the PSF photometry can be estimated with the results from the artificial star test (Section.\,\ref{artStarErr}) and is at the position of the TRGB $I_{psf,error}=0.08$\,mag.  
The uncertainty in the TRGB calibration provided by \citet{2007ApJ...661..815R} together with the contribution of the uncertainty in the metallicity is $I_{calib,error}=0.02$\,mag.
The uncertainty of the Sobel edge detection is given by the bin width and is $I_{Sobel,error}=0.05$\,mag. {Additionally, \citet{2018ApJ...852...60J} argue that  ``due to the
inability to distinguish the location of the peak within the set
binning strategy" 50\% of the bin width should be added to the error budget, corresponding to 0.025\,mag in our analysis.} Overall we obtain a total uncertainty of $I_{err}=0.10$\,mag at the distance of the dwarf galaxies, corresponding to a distance uncertainty of 0.24\,Mpc.} 

\begin{table}[H]
\caption{Properties of the two confirmed dwarf galaxies.}
\centering                          
\begin{tabular}{l c c }        
\hline\hline                 
 & dw1335-29 & dw1340-30 \\    
\hline      \\[-2mm]                  
RA (J2000)& 13:35:46 & 13:40:19\\
DEC (J2000)& $-$29:42:28 & $-$30:21:31\\
$I_{TRGB}$ (mag) & 24.43$\pm 0.10$ & 24.43$\pm 0.10$  \\
$M_I^{TRGB}$ (mag) & $-4.08\pm0.02$ & $-4.09\pm0.02$\\
$(m-M)_0$ (mag) &  28.51$\pm0.13$ & 28.52$\pm0.03$ \\
Distance (Mpc) & 5.03$\pm0.24$ &  5.06$\pm0.24$\\
$A_V, A_I$ (mag)& 0.125, 0.069& 0.158, 0.087\\
$[$Fe/H$]$ (dex) & $-1.79\pm0.4$ & $-2.27\pm0.4$\\
$M_{V}$ (mag) & $-10.25\pm0.3$ & $-10.79\pm0.3$ \\
$L_{V}$ ($10^6$ M$_\odot$) & $1.08\pm0.3$ & $1.77\pm0.3$ \\
$r_{eff,r}$ (pc) & $493\pm32$ & $414\pm32$ \\
\hline
\end{tabular}
\label{properties} 
\end{table}

\section{Discussion}
In this section we compare our photometry and results with previous work and discuss the spatial distribution of the new M\,83 satellites.

\subsection{Comparison to HST distance of dw1335-29}
\citet{2017MNRAS.465.5026C} independently discovered the dwarf galaxy dw1335-29 at the edge of an HST image taken as part of the GHOSTS survey \citep{2011ApJS..195...18R}. As the galaxy was only partially covered by the CCD chip they conducted follow-up observations using the VLT/VIMOS instrument to measure its total brightness and structural parameters. They also reported a TRGB distance based on HST photometry of 5.01$^{+0.74}_{-0.22}$\,Mpc, which is in excellent agreement with our value of $5.03\pm0.24$\,Mpc.  This agreement demonstrates that high quality images from the VLT can successfully be used to measure TRGB distances out to $\sim 5-6$\,Mpc. {In Fig.\,\ref{isoBootstrap} we overlay the HST photometry (kindly provided by Andreia Carrillo) on our VLT photometry. It is noteworthy that our estimate of the metallicity is slightly different to that of \citet{2017MNRAS.465.5026C} ([Fe/H]=-1.8\,dex opposed to [Fe/H]=-1.3\,dex){, although they are consistent within the measurement errors}. We converted the HST $F606W$ and $F814W$ photometry to $VI$ using a formula provided by Rejkuba et al. (paper in preparation). {Using the the metallicity of [Fe/H]=-1.3\,dex provided by \citet{2017MNRAS.465.5026C} we estimate a TRGB distance of 4.87$\pm0.24$\,Mpc, still well within the error estimate.}}

\subsection{Tidal features?}
{The dwarfs are close to M\,83 and are potentially tidally disturbed. In Fig\,\ref{colormap} we plotted the star map between I=24.0\,mag and I=25.5\,mag. Additionally, we color code the stars corresponding to a mask around the best fitting isochrone of the respective dwarf galaxy in red, i.e. the expected RGB stars. In the field around the galaxies we find no significant stellar overdensity. However, within the galaxy dw1340-30 there is a small lopsidedness of RGB stars in the direction South-East. Re-estimating the center with only the RGB stars gives a deviation from the previous established center of 3.2$\arcsec$. Interestingly, this lopsidedness is aligned with the connection line between M\,83 (to the North-West) and dw1340-30.}

\subsection{Spatial distribution}
\begin{figure}[ht]
\includegraphics[width=9.5cm]{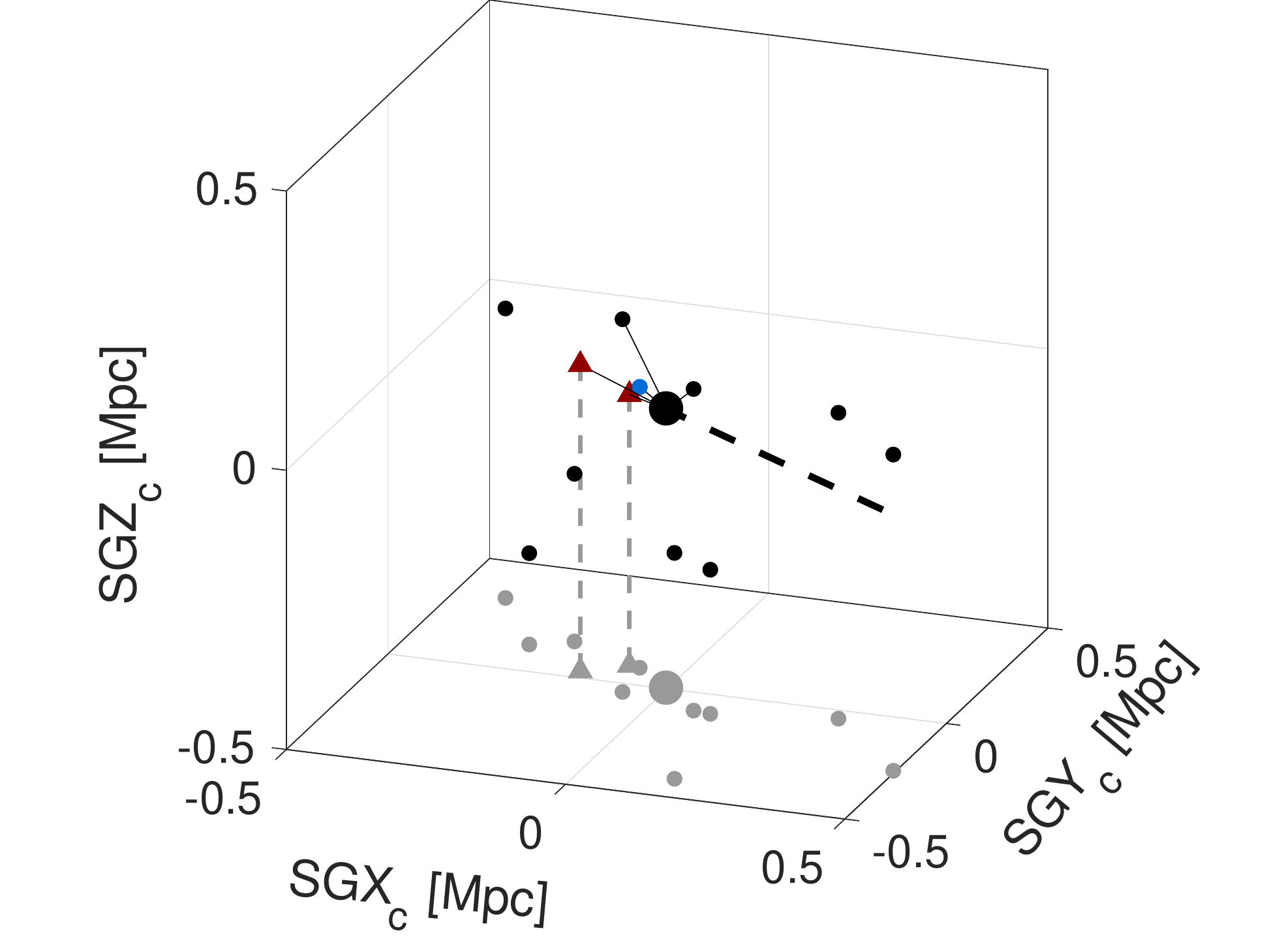}\\
\includegraphics[width=4.3cm]{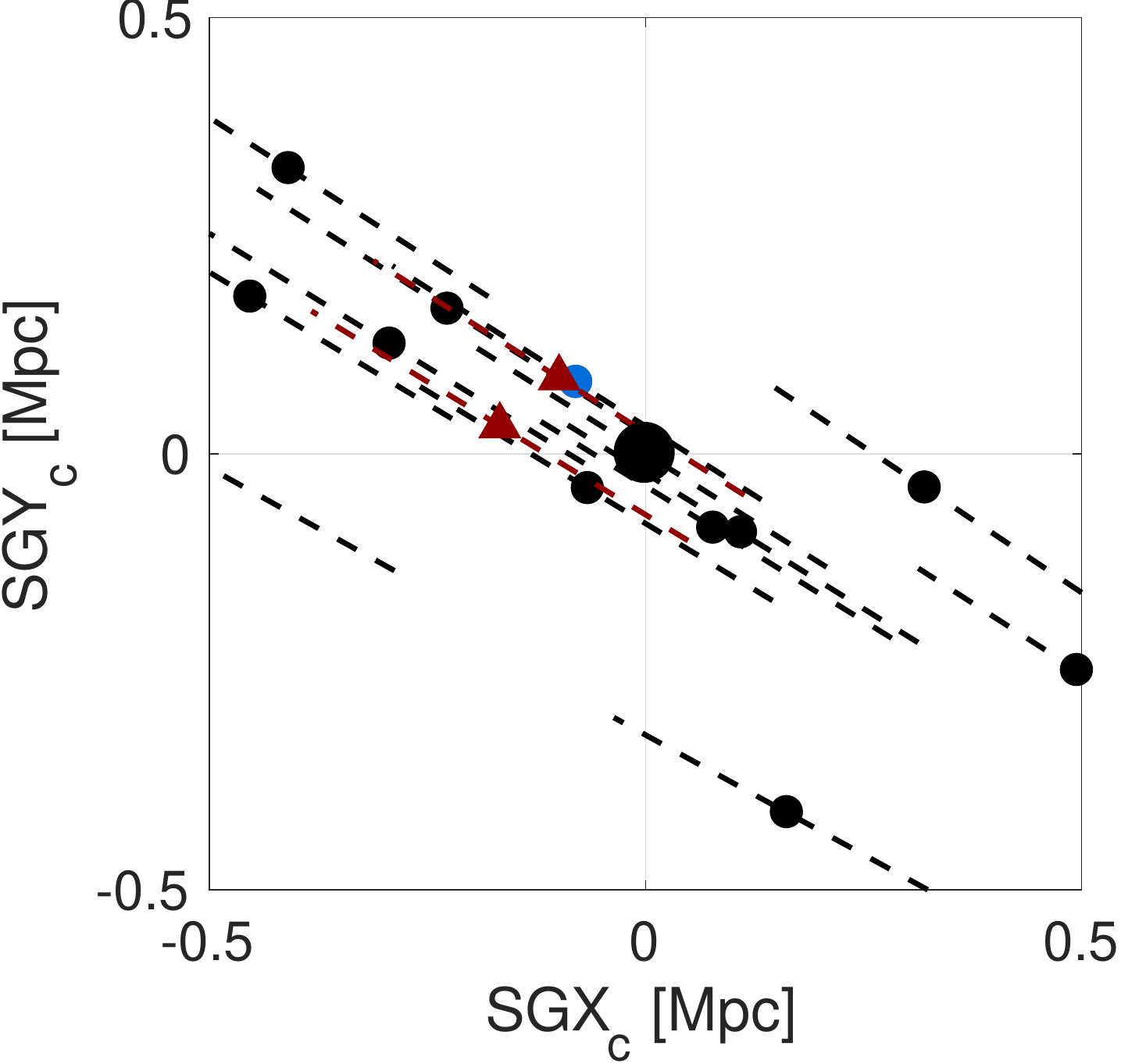}
\includegraphics[width=4.3cm]{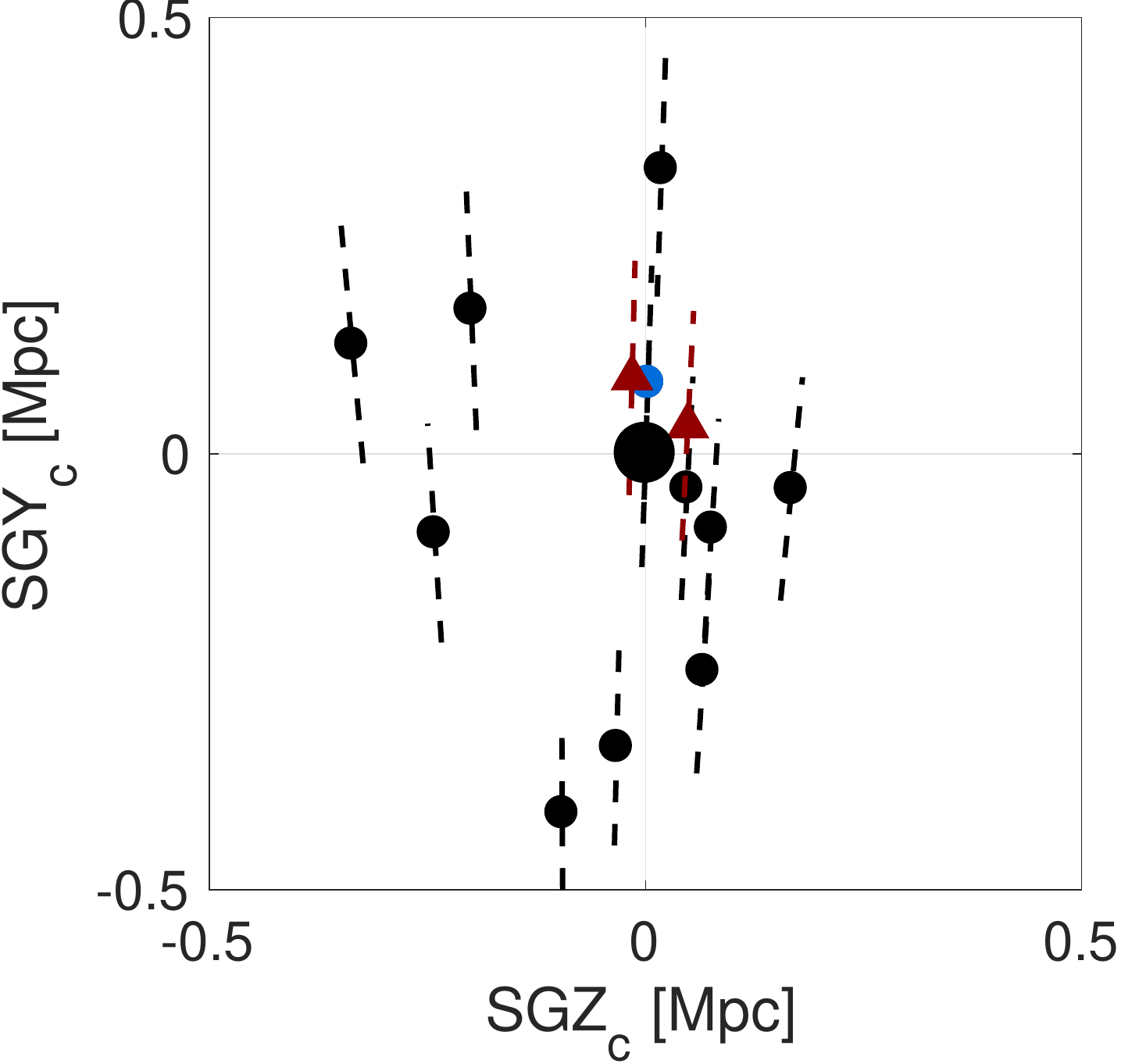}\\
\includegraphics[width=4.3cm]{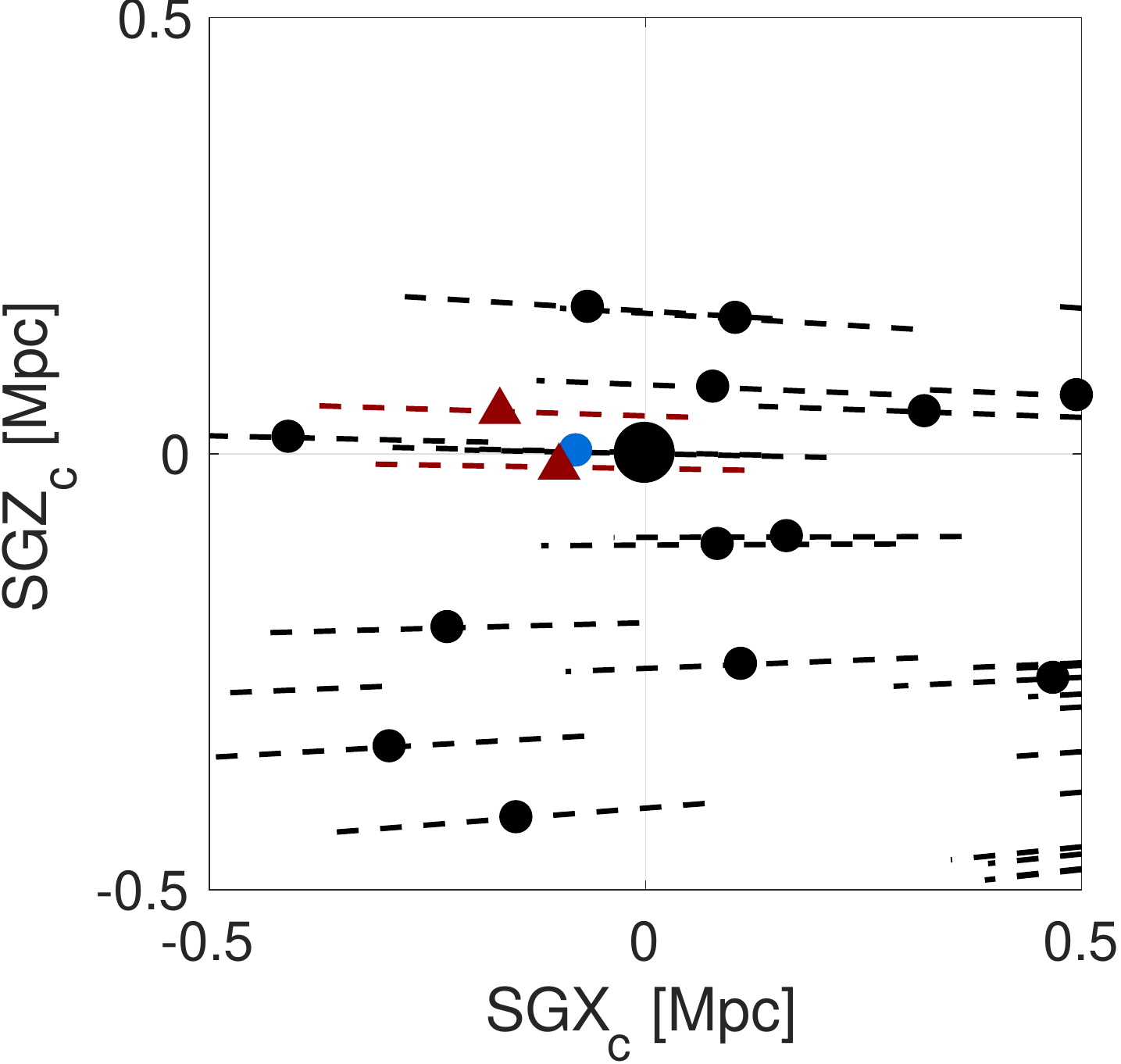}

\caption{{Distribution of the M\,83 subgroup in supergalactic coordinates, centered at M\,83. Data taken from to the LV catalog. The big black dot is M\,83, the small dots are dwarf galaxies. {Top:} satellites within the virial radius (210\,kpc) are connected to M\,83 with a thin black line. The dwarfs are projected onto the galactic plane (SGX-SGY plane). The position of the two confirmed dwarf galaxies (red triangles) are indicated with the gray dashed line. {The position of KK208 is color-coded in blue.} The black dashed line corresponds to our line-of-sight to M\,83. 
{Bottom: the projections onto the SGX-SGY, SGZ-SGY, and SGX-SGZ planes, respectively, together with their 5\% distance uncertainty (dashed lines).}}}
\label{dist}
\end{figure}

\begin{figure*}[ht]
\centering
\includegraphics[width=18cm]{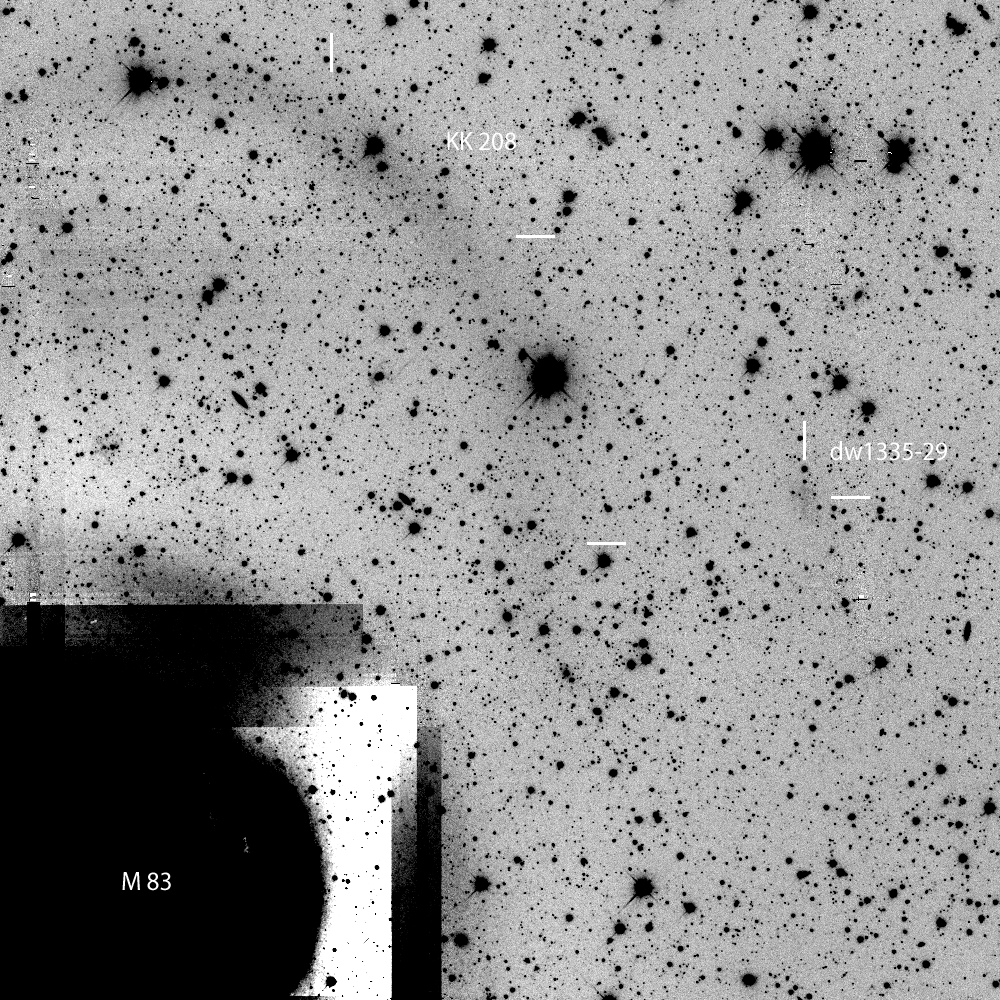}\\
\caption{{DECam field in $r$  around M\,83  \citep{2015A&A...583A..79M}, containing the tidally disrupted dwarf KK208 and dw1335-29. In the bottom left corner is M\,83.}}
\label{decam}
\end{figure*}

{Both dw1335-29 and dw1340-30 lie inside the virial radius of M\,83 {(210\,kpc; \citealt{2015A&A...583A..79M})} with a 3 dimensional separation of 130\,kpc and 180\,kpc to M\,83, respectively, making them satellites of the M\,83 subgroup {and members of the Centaurus group}. 

{\citet{2017MNRAS.465.5026C} found in dw1335-29 a relatively small but significant number of blue stars that are statistically consistent with being young upper main sequence members of the dwarf resulting from a constant star formation history. Hence dw1335-29 is either a dwarf irregular or transition type dwarf. \citet{2017MNRAS.465.5026C} noted that the presence of young stars is puzzling given the close proximity with only $\sim$26\,kpc projected distance to M\,83, and provided to possible explanations:} either the real separation is larger than the projected or M\,83 is extremely inefficient in quenching star formation. The former idea is supported by the lack of tidal features within the dwarf galaxy{, and consistent with our estimate of the 3D separation of 130\,kpc}. We further note that if there is ongoing star formation there must be gas. However, none of the candidates were detected in HI (with an upper limit of $M_{HI}< 8.5 \times 10^6$\,$M_\odot$, \citealp{2017A&A...597A...7M}).

In Fig.\,\ref{dist} we present the 3 dimensional distribution of the M\,83 subgroup in supergalactic coordinates. 
Noteworthy is the physical separation between dw1335-29 and KK208 of only 28\,kpc. This is indeed remarkably close. KK208 is a peculiar object, as it lacks a well defined center and is stretched over 10 arc minutes (corresponding to 15\,kpc at a distance of 5\,Mpc). It was resolved into individual stars with HST  \citep{2002A&A...385...21K}. \citet{2002A&A...385...21K}. The authors suggest that this object is tidally disrupted by M\,83 similar as the Sagittarius galaxy by the Milky Way \citep{1994Natur.370..194I}. {In} Fig.\,\ref{decam} {we show a DECam image that includes} dw1335-29, KK208, and M\,83. We note that \citet{2017MNRAS.465.5026C} refer to KK208 as a tidal stream. We follow the same conclusion that there is no visible connection between the two objects.

As there is strong evidence that the Cen\,A subgroup hosts a plane of satellites \citep{2015ApJ...802L..25T,2016A&A...595A.119M} one may wonder if such an arrangement is found also in the M\,83 subgroup. However, there are two major caveats {that make it rather} difficult to answer this question -- the virial radius of M\,83 is only 210\,kpc compared to $\approx 300$\,kpc of Cen\,A, meaning that a plane-of-satellites around M\,83 would be smaller compared to Cen\,A's plane ($rms$ major axis length of 309\,kpc \citealt{2016A&A...595A.119M}), and M\,83 is 1.2\,Mpc farther away, systematically increasing the error in the distance measurements (around 40 percent in absolute distance uncertainties). At the distance of M\,83, a typical distance uncertainty of $\pm$0.3\,Mpc is already covering the full virial radius. In Fig.\,\ref{dist} we projected the dwarf satellites onto the galactic plane. 
{ 
Using a singular value decomposition as described in \citet{2016A&A...595A.119M} we fit a plane through the M\,83 subgroup and estimate a $rms$ thickness of 20.4\,kpc, considering all 6 galaxies (KK\,208, M\,83, NGC\,5264, KK\,218, dw\,1340-30, and dw\,1335-29) within 210\,kpc (= the virial radius of M\,83 \citet{2015A&A...583A..79M}) and $rms=55$\,kpc for galaxies within 600\,kpc (while removing HIDEEP\,J1337-33 from the analysis as this dwarf is $\approx$\,300\,kpc away from the estimated plane).
In comparison: the Milky Way plane has a $rms$ thickness of 19.9\,kpc , the Andromeda plane a $rms$ thickness of 13.6\,kpc  \citep{2013MNRAS.435.1928P}, the Cen\,A plane a $rms$ thickness of 69\,kpc \citep{2016A&A...595A.119M}, and the plane around M\,101 a $rms$ thickness of 46.0\,kpc \citep{2017A&A...602A.119M}.}
}

\section{Summary and conclusions}
We have resolved the two dwarf galaxy candidates dw1335-29 and dw1340-30 \citep{2015A&A...583A..79M} into individual stars, using the FORS2 instrument mounted at the VLT. With these deep images we derived their distances and mean metallicities and confirmed them to be members of the M\,83 subgroup in the Centaurus group. {The third dwarf candidate dw1325-33 could not be resolved into individual stars, due to} {increased sky brightness preventing sufficiently deep photometry.} When comparing to similar group studies, i.e. the M\,81 group \citep{2009AJ....137.3009C,2014ApJ...795L..35C} or the M\,101 group \citep{2014ApJ...787L..37M,2017ApJ...837..136D}, a confirmation rate of $\sim$67\% is expected and achieved. 

Future follow-up observations for the numerous remaining dwarf galaxy candidates are necessary to create a complete picture of the Centaurus group. As most of the remaining dwarf galaxy candidates will be associated to the closer Cen\,A subgroup, {resolving their individual} red giant branch stars will be easier to achieve {assuming a similar} observation strategy we used with the VLT. {At the Centaurus A distance the RGB tip magnitude is approximately 0.4\,mag brighter at $M_{I,TRGB} = 24.05 \pm 0.05$ \citep{2005ApJ...631..262R}. The other members of this subgroup are expected at a similar distance, implying that their RGB tip is} well within our detection and completeness limits.

Why is it important to aim for a well sampled census of the Centaurus group dwarfs? Apart from studying the star formation history and the metallicity distribution of the group \citep{2006A&A...448..983R, 2010A&A...516A..85C,2011A&A...530A..58C,2011A&A...530A..59C,2012A&A...541A.131C}, which is in itself an interesting topic, the Centaurus group provides an excellent testbed to study cosmological model predictions, e.g.~the phase-space correlation of the Cen\,A satellite galaxies \citep{2018Sci...359..534M}.
Such results rest fundamentally on the number of accurate distance measurements and can be affected by small number statistics. However, if the co-rotating planar structure is true, the addition of new data will increase its significance and further support this finding. It is therefore a crucial task to continue measuring distances for as {many} candidates as possible.

\begin{acknowledgements}
OM is grateful to the Swiss National Science Foundation for financial support. HJ acknowledges the support of the Australian Research Council through Discovery project DP150100862. {We thank Andreia Carrillo for providing the HST photometry for dw1335-29 {and the referee for helpful comments, which improved the clarity of the paper}.}
\end{acknowledgements}


\bibliographystyle{aa}
\bibliography{aanda}

\end{document}